# A Practical Implementation of Customized Scrum-Based Agile Framework in Aerospace Software Development Under DO-178C Constraints

Malik Muhammad Umer

College of Computing, Georgia Institute of Technology, Atlanta, GA 30332 USA, malikumerpak@gmail.com

The increasing complexity of aerospace systems necessitates software development processes that harmonize agility with stringent safety and certification obligations. This study presents an empirically validated Scrum-based Agile framework tailored for the development of DO-178C–compliant safety-critical aerospace software. The framework introduces systematic adaptations to conventional Scrum roles, artifacts, and events to fulfill certification, verification, and independence objectives mandated under DO-178C. Key enhancements include a multi-disciplinary product ownership model, dual compliance–functional acceptance criteria, independent testing and documentation teams, and integrated certification liaisons. The proposed framework was evaluated through two comparable aerospace projects—one employing the customized Agile process and the other following a traditional Waterfall model—to assess impacts on efficiency, quality, and compliance. Results demonstrated measurable improvements, including 76% reduction in Total Effort per Requirement, 75% faster Defect Detection Rate, 78% faster Defect Resolution Time, and over 50% lower Defect Density, all achieved while maintaining full compliance with DO-178C Design Assurance Level (DAL) A objectives. These findings confirm that Agile practices and regulatory compliance can coexist effectively when supported by disciplined process tailoring and proactive engagement with certification authorities. The study also identifies notable challenges, including increased Verification and Validation (V&V) effort due to recurring activities across Sprints and frequent code refactoring inherent to iterative development. Nonetheless, it highlights significant opportunities for further enhancement through greater workflow automation, Continuous Integration and Delivery (CI/CD), and automated documentation, verification, and configuration management. Future research is recommended to extend validation of this framework across the aerospace sector and other safety-critical industries governed by similarly rigorous certification standards.

## 1 INTRODUCTION

Software development in the aerospace industry is subject to rigorous regulation under the DO-178C standard [1], which establishes the framework for ensuring the reliability, safety, and certifiability of airborne software systems. Officially titled "Software Considerations in Airborne Systems and Equipment Certification," DO-178C outlines a comprehensive set of objectives and associated activities across various lifecycle processes, including planning, design, development, testing, verification, and configuration management. These objectives are determined by the safety criticality level of the software. For instance, Flight Control Computer software is typically classified as Level A, the highest level of criticality, since its failure could result in catastrophic consequences, including potential loss of life and aircraft. Consequently, software at this level must undergo the most rigorous development and verification to satisfy all mandated objectives.

Aerospace organizations conform to DO-178C guidelines not only to assure quality and safety but also to produce the necessary evidence for regulatory certification [2], [3]. While the standard specifies what must be achieved, it does not prescribe how-that is, it allows flexibility in the choice of software development models. Whether employing traditional lifecycle models such as Waterfall [21] or adopting Agile-based frameworks like Scrum [23], the selection depends on project-specific attributes-including system complexity, requirements stability, reuse of legacy software, development strategy, and hardware availability. Historically, the Waterfall model has been predominantly used in aerospace to maintain compliance and generate safety-critical artifacts required for successful certification [4], [5]. This preference stems from the model's compatibility with DO-178C objectives, which emphasize thorough documentation, detailed planning, and upfront requirement specification. Additionally, the Waterfall and V-model [22] integrate effectively with broader systems engineering and hardware development workflows. In aerospace projects, requirements are typically well understood and are expected to be defined early in the lifecycle. Moreover, limited customer interaction in certain contexts further reinforces the adoption of these traditional models [6].

To remain competitive with the evolving software industry, the aerospace sector is increasingly under pressure to adopt Agile methodologies that enable more rapid and flexible responses to changing customer needs. A survey conducted in 2023 among software industry professionals revealed that 71% of the respondents reported the adoption of Agile methodologies in their development processes [7]. Consequently, both industry practitioners and researchers have begun to investigate the applicability of Agile practices in this domain. Despite the traditionally exhaustive upfront requirements analysis phase characteristic of aerospace projects, it is now widely recognized that requirement changes are often inevitable during software development. However, within the constraints of the Waterfall model, incorporating such changes during the implementation phase is typically difficult and costly [5], [8].

The motivation to transition toward Agile practices is also driven by the desire to leverage key advantages-such as early user feedback, timely identification of defects or misalignments with stakeholder expectations, and the reduction of late-stage rework-which can be expensive and time-consuming. Importantly, the adoption of Agile in aerospace is not intended to circumvent the need for rigorous documentation or planning-critical elements under standards such as DO-178C-but rather to enhance software quality and, by extension, system safety. Moreover, Agile methodologies offer several secondary benefits-including improved development efficiency, reduced time to market, enhanced customer satisfaction, and lower overall project costs. Agile is a highly disciplined approach to software and systems engineering that emphasizes building quality into the product from the outset, supported by continuous verification and validation throughout the development lifecycle [9].

As the software industry continues to evolve rapidly, aerospace organizations are increasingly exploring the adoption of Agile methodologies to enhance flexibility, improve responsiveness to change, and reduce development costs. Agile practices offer several advantages-including early user feedback, faster discovery of defects and inconsistencies, improved productivity, and shorter time to market. However, the application of Agile in safety-critical domains such as aerospace remains challenging due to the stringent regulatory and certification requirements enforced by standards like DO-178C. DO-178C mandates comprehensive documentation, strict traceability, and independent verification and validation, all of which necessitate significant modifications to conventional Agile practices. The structured and plan-driven nature of aerospace software development, particularly when integrated with hardware and systems engineering efforts over multi-year timelines, poses difficulties for adopting Agile's inherently flexible and iterative approach.

Critics argue that the fundamental assumptions underlying Agile, such as evolving requirements, minimal documentation, and frequent customer collaboration, are often incompatible with the characteristics of aerospace projects [10], [11]. For instance, incorporating requirement changes during the development phase is particularly costly in the traditional Waterfall model, which aerospace companies have historically relied upon for its alignment with certification standards [5], [7]. Agile's emphasis on minimizing documentation, including the bi-directional traceability, conflicts with the need for extensive evidence to demonstrate compliance. Moreover, Agile methodologies are less suited to hardware-dependent environments where prototyping is time-consuming and expensive [13]. The prescriptive processes defined by DO-178C further constrain the development team's ability to inspect and adapt workflows in real time, limiting one of Agile's core strengths-continuous process improvement [12], [14].

Despite these challenges, several studies suggest that Agile practices can be adapted for use in safety-critical systems. Notander et al. [12], for example, conclude that Agile methods are not inherently incompatible with safety-critical standards but require careful tailoring to align with regulatory constraints. The concept of method adaptation has been documented across various software domains [15], [16], yet there is limited empirical evidence in the aerospace context regarding which adaptations are most effective or how they are implemented in practice. Existing literature indicates that although some Agile practices have shown promise, no single approach has yet been fully validated in real-world aerospace projects, particularly for compliance with DO-178C Level A-the highest and most rigorous certification level [5], [17], [18], [19], [20]. Agile's iterative development model often struggles to maintain the level of traceability, formal verification, and documentation demanded by aerospace standards.

Therefore, while Agile methodologies offer considerable potential for enhancing development efficiency and software quality in the aerospace domain, their effective application requires deliberate adaptation and integration with existing regulatory frameworks. This research addresses that challenge by proposing and validating a customized Scrum-based Agile development process that combines the flexibility and iterative nature of Agile with the rigor and structure mandated by the DO-178C standard. The proposed model is based on the Scrum framework, selected due to its widespread adoption across industries. According to the 17th State of Agile Report (2023), 63% of Agile practitioners utilize Scrum as their primary methodology [7]. The customized Scrum-based process was practically implemented within an aerospace organization as part of a real-world industrial project involving the development of a DO-178C Level A safety-critical avionics software system, which represents the highest level of rigor in airborne software certification. This study presents a comprehensive overview of the steps taken to adapt and incorporate the Scrum framework in the context of safety-critical aerospace software development. To guide the development and evaluation of the proposed approach, four primary research questions were defined:

- **RQ1:** To what extent does the proposed Scrum-based Agile model align with the core values of Agile and Scrum?
- **RQ2:** Can compliance with DO-178C objectives be achieved within the proposed Scrum framework?
- **RQ3:** How effectively does the proposed approach adequately address the requirements of certification authorities?
- **RQ4:** Does the proposed approach introduce measurable efficiencies into existing processes?

The remainder of this paper is organized as follows: Section 2 presents a comprehensive overview of Agile methodologies, with a particular focus on the Scrum framework. Section 3 outlines the objectives, activities, and evidence artifacts required by the DO-178C standard for software certification in safety-critical aerospace systems. Section 4 provides background information on the company, including its existing development processes prior to the adoption of Scrum, thereby establishing the context for the proposed Agile transformation. Section 5 details the practical implementation of the customized Scrum framework within the company. Section 6 evaluates the proposed approach in detail, highlighting both the realized benefits and encountered challenges, while also addressing the defined research questions. Section 7 reviews related work in the field, situating the present study within the broader research landscape.

Section 8 discusses the limitations of the current study and outlines potential directions for future research. Finally, Section 9 summarizes the key conclusions drawn from the findings of this research.

## 2 OVERVIEW OF AGILE METHODS AND SCRUM FRAMEWORK

This section presents a foundational overview of Agile software development, highlights the key features of the Scrum framework, and discusses the potential benefits and challenges associated with implementing Agile methodologies within the aerospace industry.

### 2.1 Agile Software Development

Agile software development emerged in the late 1990s as a response to the limitations and shortcomings of traditional plan-driven methodologies, such as the Waterfall model [24], [25], [26], [27] and the Rational Unified Process (RUP) [28], [29], which struggled to accommodate the dynamic and rapidly evolving nature of software requirements. One of the primary criticisms of these conventional approaches is the significant lag between the initiation of development and software delivery, which often fails to keep pace with the changing demands of the problem domain [29], [30], [31], [32]. For instance, typical iteration cycles in RUP span six to twelve months, during which time project requirements or available technologies may shift significantly. In contrast, advocates of Agile methodologies [24], [32] promote a development model centered on continuous progress review and close, iterative collaboration with customers. As Schwaber and Beedle [30] and Abrahamsson et al. [32] explain, this philosophy is rooted in empirical process control, where processes are not rigidly predefined but are instead adapted incrementally based on ongoing observation and feedback. Consequently, Agile teams begin development with a high-level understanding of project goals but deliberately limit detailed planning to near-term tasks to retain flexibility and responsiveness.

The Agile Manifesto [24] articulates a foundational shift from traditional, plan-driven software development approaches to a more adaptive, human-centric paradigm. It emphasizes four core values: prioritizing individuals and interactions over processes and tools, working software over comprehensive documentation, customer collaboration over contract negotiation, and responding to change over strictly adhering to a predefined plan. These values collectively support a flexible and iterative development process that fosters continuous feedback and rapid delivery of functional software, thereby enabling development teams to better accommodate evolving user needs and changing requirements. Although formal documentation, structured planning, and tool usage remain relevant, Agile approaches emphasize dynamic communication, early value delivery, and adaptive responsiveness, qualities particularly advantageous in volatile or complex project environments. These core values serve as the philosophical underpinning for widely adopted Agile methodologies such as Scrum [23], Kanban [34], and Extreme Programming (XP) [33], as well as scaling frameworks including the Scaled Agile Framework (SAFe) [35], Scrum@Scale (S@S) [36], and Disciplined Agile Delivery (DAD) [37]. A distinguishing characteristic shared by these methodologies is their iterative and concurrent nature. Development is typically carried out in short time-boxed cycles, commonly two to three weeks in duration, though some may span only a single day. These iterations are frequently concluded with deliverables presented to stakeholders for immediate feedback and review. In contrast to traditional plan-driven models, Agile encourages the concurrent execution of development activities such as requirements elicitation, design, coding, and testing within each iteration, thereby enhancing efficiency and adaptability throughout the project lifecycle.

## 2.2 Scrum Framework

To effectively understand the proposed Agile development approach and its implementation within the aerospace industry, it is essential to first provide a comprehensive overview of the Scrum framework. A detailed exposition of the complete Scrum methodology is available in the official Scrum Guide [23].

Scrum is an Agile process framework that facilitates iterative and incremental software development through a defined set of roles, events, and artifacts. It emphasizes adaptive planning, early and frequent delivery of working software, and continuous improvement, all while fostering close collaboration, shared responsibility, and responsiveness to stakeholder feedback. Originally developed by Schwaber and Sutherland [23], Scrum has gained widespread adoption across various industries due to its clarity, adaptability, and proven effectiveness in managing complex software projects.

The Scrum process is structured around fixed-length development cycles known as Sprints, typically lasting two to four weeks. Schwaber et al. [23] define Sprints as time-boxed iterations of no more than one month, providing a consistent rhythm for the team's development activities. Prior to each Sprint, the project requirements are organized into a Product Backlog, a dynamic and prioritized list encompassing all desired features, enhancements, and fixes for the product. A key prerequisite for initiating a Sprint is the availability of well-defined Product Backlog Items (PBIs), as multiple iterations are generally required to address the complete backlog. Each Sprint commences with a Sprint Planning session, during which the Scrum Team selects a subset of high-priority PBIs and formulates a corresponding Sprint Goal. The selected PBIs are further refined into smaller, actionable tasks known as Sprint Backlog Items. These collectively form the Sprint Backlog, which outlines the team's executable plan for upcoming iteration. Sprint culminates in the delivery of a potentially shippable product increment, providing tangible progress and enabling frequent validation against customer expectations.

Daily progress is monitored through a brief, time-boxed meeting known as the Daily Scrum, during which each team member briefly reports progress, plans and any issues that have arisen, which enhances transparency, fosters synchronization, and enables the team to adjust plans as needed. At the conclusion of each Sprint, the team conducts a Sprint Review to demonstrate potentially shippable product increment to stakeholders and gather feedback. This is followed by Sprint Retrospective, during which the team reflects on Sprint process and identifies areas for improvement.

Scrum delineates three key roles that structure its operational framework. The Product Owner is responsible for maximizing the value of the product by managing the Product Backlog and serving as the primary liaison between the development team and stakeholders. The Scrum Master acts as a facilitator, ensuring the proper implementation of Scrum practices, removing obstacles that hinder progress, and fostering an environment conducive to team effectiveness. The Development Team comprises a cross-functional group of professionals tasked with delivering potentially shippable product increments at the end of each Sprint. According to Schwaber and Beedle [30] and Lei et al. [39], Scrum is particularly effective for small, collaborative teams typically consisting of three to nine members.

Scrum is supported by three core artifacts-the Product Backlog, the Sprint Backlog, and the Increment-which promote transparency and enable empirical process control. The methodology is grounded in three foundational pillars: transparency, inspection, and adaptation. These principles ensure that decisions are made based on observable outcomes rather than assumptions, fostering responsiveness to change and continuous improvement. By combining structured roles with iterative delivery cycles, Scrum facilitates the development of high-quality software, accommodates evolving requirements, and supports sustained team performance within dynamic project environments.

Evolution of Scrum practice has been analyzed in terms of three distinct modes of Sprint execution, commonly referred to as Type A, Type B, and Type C Sprints, as illustrated in Figure 1. This categorization, introduced by Jeff Sutherland [74], reflects how Scrum teams structure their workflow, particularly regarding the sequencing and overlap of activities across successive Sprints. The typology is grounded in earlier framework by Takeuchi and Nonaka [75], who described product development processes as varying between sequential, partially overlapping, and fully overlapping phases.

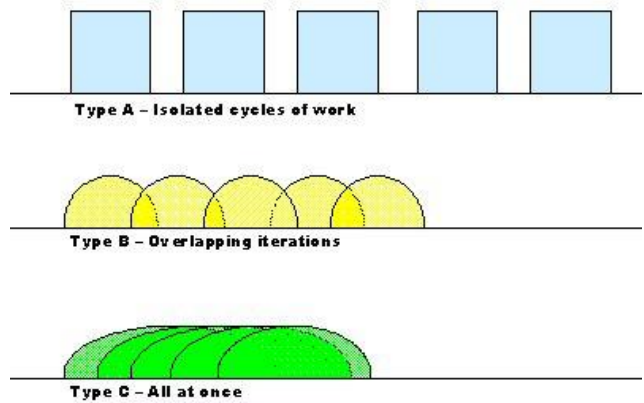

Figure 1: Scrum Type A, B, and C, adapted from [74].

Type A Sprints represent the most traditional form, in which all work related to a Sprint is executed strictly within the timebox of that Sprint. Once the Sprint concludes, the next Sprint begins only after planning and preparation activities are complete. This model ensures clear boundaries between Sprints, simplifying management and scope control, but it may introduce idle time between iterations. Type B Sprints incorporate preparatory work for the subsequent Sprint within the current iteration. By allowing partial overlap, this approach reduces transition gaps and facilitates smoother continuity between Sprints. However, it also increases planning complexity and requires teams to manage the potential risk of scope creep or misalignment of priorities. Type C Sprints represent the most advanced form, in which multiple Sprint activities overlap significantly, enabling a near-continuous flow of work. In this model, definition, development, and testing phases are conducted concurrently across iterations. Type C is often associated with mature Scrum teams employing continuous integration and delivery practices. While this model achieves higher efficiency and shorter cycle times, it demands robust automation, disciplined backlog management, and advanced team coordination.

In practice, organizations may adopt different Sprint types depending on their maturity, technical infrastructure, and project requirements. Type A is suitable for teams in early stages of Scrum adoption, Type B for intermediate teams seeking improved flow, and Type C for highly mature teams aligned with DevOps and continuous delivery environments.

**2.3 Agile Application in Aerospace**

Implementing Agile methodologies within the aerospace industry presents considerable challenges, primarily due to the stringent regulatory and certification requirements inherent in safety-critical domains. Standards such as DO-178C [5], which govern the development of airborne software, demand rigorous documentation, traceability, and formal verification processes. These requirements often stand in contrast to the core principles of Agile, which emphasize flexibility, iterative development, and minimal documentation. Agile approaches traditionally deprioritize traceability—focusing on delivering functional software over documenting the process by which it is developed—thus conflicting with the validation and assurance demands imposed by safety-critical standards. In particular, conventional Agile practices do not inherently align with the V-Model (Figure 2), a fundamental framework in safety-critical software development that mandates systematic validation and verification activities at each development stage. The iterative and adaptive nature of Agile complicates the enforcement of strict traceability and formal verification required for compliance with standards such as DO-178C, particularly at the highest assurance level (Level A).

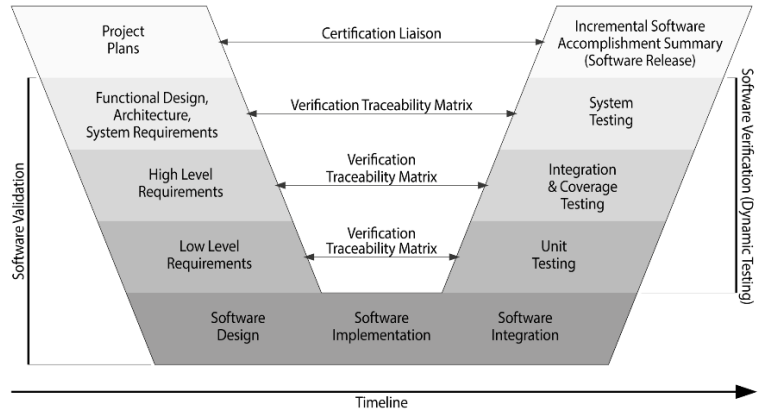

Figure 2: DO-178C V-Model software development process, adapted from [1], and [100].

The adoption of Agile practices in safety-critical software development has shown promising results, despite the constraints posed by rigorous regulatory frameworks. VanderLeest et al. [8] conducted pilot studies applying various Agile techniques within DO-178B-compliant projects and reported positive outcomes. Notably, practices such as pair programming, fixed-length iterations, test-driven development (TDD), and continuous integration (CI) were identified as particularly effective in addressing some of the challenges inherent to aerospace software development, even within the rigid structure of the DO-178B standard. However, their study lacked a detailed exposition of the implementation strategies and the specific outcomes achieved. The authors themselves acknowledged the need for further empirical validation and broader demonstrations.

Complementing these findings, a systematic literature review by Kasauli et al. [62] highlighted a range of benefits associated with Agile adoption in the development of safety-critical systems. The review synthesized key advantages frequently cited in the literature, including enhanced stakeholder engagement, cost reduction, improved software quality, more effective utilization of available information, strengthened safety culture, increased potential for component reuse, better management of evolving requirements, improved requirement prioritization, alignment of functional and safety requirements, and the development of more robust test cases.

Marques et al. and Vuori have emphasized that Agile methodologies are expected to enhance project control, enable earlier delivery of value, and support adaptability to evolving requirements more effectively than traditional lifecycle models. Key drivers for adopting Agile approaches include the delivery of early partial products, improved predictability, better alignment with customer expectations, reduced system complexity, and earlier risk identification and mitigation [40], [41]. However, the integration of Agile within safety-critical domains continues to face significant hurdles. Late-stage requirement changes and documentation management are consistently cited as major challenges [5], [42]. In particular, the documentation-heavy emphasis of standards such as DO-178C can conflict with the Agile principle of minimizing non-essential documentation. Certification authorities are generally unwilling to accept reductions in documentation related to software requirements, design, or verification activities, all of which are necessary to demonstrate system quality and safety compliance [42], [43], [44], [45].

Despite this, Carbone et al. argue that documentation need not be a barrier, asserting that Agile methods-when properly applied-can fulfill customer requirements, including those related to safety-critical documentation. They recommend focusing documentation efforts only on essential aspects that contribute directly to safety or traceability [42]. A review of related literature from other safety-critical industries reinforces this viewpoint, highlighting two recurring concerns: effective management of evolving requirements and the need for regulatory-compliant documentation [46], [47], [48], [49]. It is commonly acknowledged that while functional requirements may evolve over time, safety requirements tend to remain stable. Consequently, researchers recommend separating functional and safety requirements in the product backlog [48], [50]. To address these challenges, Myklebust et al. [51] propose the use of "safety stories"-structured in a manner analogous to user stories-to encapsulate safety requirements within Agile frameworks. Others suggest maintaining separate documentation for functional requirements and hazard analysis during the initial phases of a project and integrating them at a later stage [52]. In safety-critical contexts, Agile methods must support extensive documentation capable of justifying safety arguments to certification authorities [53], [54].

Early transition studies by Rodrigues et al. and VanderLeest et al. demonstrated that it is feasible to shift from traditional Waterfall approaches to Agile within safety-critical domains without abrupt disruption to existing processes. Instead, a gradual transition can be achieved through the integration of Agile practices, supported by process automation and incremental change [8], [55]. Further contributions from Ribeiro et al., Baron et al., and Marsden et al. provide deeper insight into the challenges of developing aerospace software in compliance with DO-178C [5], [18], [56]. Baron et al. acknowledge that while some Agile principles may not align directly with certification requirements, they can be adapted effectively. Their work emphasizes embedding certification considerations throughout the development lifecycle, rather than treating them as post-development add-ons [56]. Building upon earlier findings, Baron et al. [19] further explored the application of Agile to reduce costs and improve quality in the development of avionics embedded software, within the DGA TA framework [57]. Their proposed hybrid framework incorporates Agile principles such as Daily Stand-up meetings, Planning Poker, Minimum Viable Product (MVP) strategies, defined roles, ticket-based project management, use of open-source tools, and off-the-shelf components. A case study demonstrates the potential for cost savings and efficiency gains. However, the approach is not without limitations. Challenges include the need for a paradigm shift in certification practices, substantial upfront investment, institutional risk aversion during audits, and a shortage of engineers with training in Agile methodologies and their application in safety-critical domains [19].

Marsden et al. conducted a detailed investigation into the compatibility and reconciliation of Agile practices with the stringent documentation requirements of ED-12C [58] and DO-178C standards in aerospace software projects [18]. Their study explored three distinct process models, all derived from Scrum principles, to enhance agility within such regulated environments: Incremental Process, Functional Process, and Agile Process. The Incremental Process is structured around successive releases, each delivering a functional increment. The Functional Process organizes work into discrete functional packages, while the Agile Process aims for finer granularity by executing the complete software development lifecycle for each individual feature or function. Within this Agile model, quality gates are aligned with finalized requirements, design, and testing procedures at the culmination of each Sprint. To address the inherent risks of transitioning between process phases, the concept of "micro-reviews" was introduced—targeted internal reviews applied to every transition of each function. However, this approach demands careful management to prevent conflicts arising from user-focused requirements or suboptimal change control mechanisms. Marsden et al. carried out four case studies across different divisions within an organization. The outcomes indicated potential benefits, but several limitations were observed, including excessive dependency on a single Product Owner (PO), duplicated verification activities, and inadequacies in the maturity of functional requirements. The Key Performance Indicators (KPIs) used-namely, the number of Problem Reports (PRs) and their resolution time within release cycles—provided only a limited basis for evaluating the effectiveness of the proposed methodologies [18].

In a related contribution, Aradhya proposed an enhanced Agile process framework specifically tailored to Scrum for use in aerospace software development [20]. This framework comprises three iterative phases: Preparation, Development, and Closure. New functionalities are developed and integrated during the Development phase, while system-level testing, final documentation, and release preparation are conducted in the Closure phase. These phases recur throughout the development lifecycle until the software product is finalized [20]. Aradhya acknowledged the necessity for further investigation, particularly with respect to establishing a comprehensive and stable software specification early in the project and maintaining up-to-date software architecture throughout development. Nevertheless, the proposed framework demonstrates potential for addressing domain-specific challenges within avionics systems [20].

Costa et al. [17] extended the work of Aradhya [20], identifying specific barriers that complicate the application of Agile methods in the development of embedded aeronautical software compliant with the DO-178C standard. They proposed adaptations to the Scrum methodology and structured their framework into four phases: Pre-game (encompassing planning, documentation, and initial architecture), Game (focused on iterative development through "Reconfigurable Sprints"), and Post-game (final architecture validation and documentation). While these phases suggest a structured approach, the "Game" phase employs the standard Scrum framework as per the official Scrum Guide [23], without evident modifications to address domain-specific needs. Furthermore, Costa et al. did not elaborate on the practical implementation of these adaptations beyond the defined phases, and the study lacks empirical validation through surveys or case studies. They concluded that expert evaluation of their proposed adaptations by DO-178C specialists is a necessary next step. Notably, their findings confirm persistent challenges in applying Scrum within safety-critical, regulated contexts, particularly regarding compliance with rigorous certification and documentation mandates [17].

Eduardo et al. [60] introduced Scrum4DO178C, an Agile process model grounded in the Scrum framework, specifically developed to address the stringent demands of aerospace software development, including safety, robustness, reliability, and integrity. This framework incorporates targeted enhancements designed to align Agile practices with the requirements of the DO-178C standard. In contrast to earlier approaches that often lack practical detail, Scrum4DO178C presents a comprehensive and empirically validated process, successfully implemented in an industrial project classified at the highest criticality level (Level A – Catastrophic). The study's outcomes indicate that the proposed model enhances overall project performance by enabling more frequent and manageable changes to requirements, reducing V&V overhead, and improving development efficiency—all while ensuring full compliance with DO-178C certification requirements.

Heager and Nielsen [63], through a series of interviews with teams involved in critical software development, identified four principal challenges to the adoption of Agile practices in safety-critical contexts. First, the use of documentation presents a significant obstacle; while Agile methodologies typically advocate for minimal documentation, compliance with certification standards necessitates comprehensive and traceable documentation, creating a tension between agility and regulatory requirements. Second, requirements engineering is complicated by the inherent uncertainty of evolving requirements, which can adversely affect developers' ability to plan and implement solutions effectively. Third, the software development lifecycle is frequently disrupted due to the need for coordination with concurrent hardware development efforts, particularly in embedded systems. As highlighted by Kasauli et al. [64], these disruptions—often stemming from dependencies on hardware teams—undermine the continuity and rhythm of software iterations. Finally, testing in an Agile context introduces additional complexity, as transitioning to iterative testing approaches demands significant modifications to established testing workflows. This shift often requires testers to acquire new skills and adapt to unfamiliar practices, posing a barrier to seamless integration. These limitations and concerns represent some of the most salient issues documented in the literature regarding the adoption of Agile methodologies in aerospace software development governed by DO-178C.

Documentation remains one of the most significant challenges in the development of safety-critical software systems [5], [47], [55]. In this context, Rodrigues et al. proposed an automated mechanism aimed at improving document processing and management within aerospace software projects [55]. Their study focuses on increasing the efficiency and reliability of documentation practices in accordance with the stringent demands of safety-critical environments. The findings indicate that automation of documentation processes is both feasible and beneficial in this domain. However, as Rodrigues et al. caution, any tools developed for such purposes must conform to the requirements of the DO-330 standard [59], which governs tool qualification for airborne systems. Achieving compliance with this standard is a complex and demanding endeavor, highlighting the need for meticulous tool design and validation.

Aerospace projects typically emphasize software development and system-level activities, each characterized by distinct workflows, deliverables, and complexities. These projects may progress through various stages of the V-Model (Figure 2), encompassing phases from system specification to integration and validation. To address the inherent challenges of managing such complexity, Eduardo et al. [60] examined strategies for refining task granularity. Their approach was operationalized within Jira [61], a widely adopted work management tool across the aerospace industry and its clientele. The use of Jira facilitated the management of safety-critical project requirements. Nevertheless, the authors encountered several challenges, including the decomposition of High-Level Requirements (HLRs) into Product Backlog Items (PBIs) that could be completed within short iteration cycles (typically one month or less); the formation of development teams capable of implementing these PBIs while fulfilling stringent verification and validation (V&V) criteria; and the decision of whether a single team could effectively manage the broad scope of activities, or if the distribution of tasks across multiple specialized teams was more appropriate.

The literature acknowledges that while certain Agile practices have demonstrated promise within aerospace and other safety-critical sectors, no single Agile methodology has been comprehensively validated and adopted in large-scale, real-world industrial projects [5], [17], [18], [19], [20]. Although several adaptations and hybrid models have been proposed, none have successfully achieved full compliance with the rigorous certification requirements outlined by standards like DO-178C. This study addresses these limitations by proposing and validating a novel process framework that enables the integration of Agile practices within the constraints of aerospace software development. It introduces a tailored set of implementation and transition steps designed to reconcile the flexibility of Agile with the structured demands of regulatory standards, thereby facilitating safe, certifiable, and responsive software development in aerospace contexts. Moreover, the study uniquely contributes to the body of knowledge by documenting the organizational experience of navigating transitional challenges—particularly employee resistance to change and the critical role of management commitment during the shift from traditional to Agile methodologies.

## 3 OVERVIEW OF DO-178C OBJECTIVES, ACTIVITIES AND DATA

DO-178C [1], formally titled "Software Considerations in Airborne Systems and Equipment Certification", is a guidance document developed by RTCA and EUROCAE to ensure that software used in airborne systems meets the rigorous safety and reliability requirements necessary for certification by aviation authorities such as the FAA (Federal Aviation Administration) and EASA (European Union Aviation Safety Agency). It supersedes its predecessor, DO-178B, by introducing improved guidance, clarifying existing objectives, and offering greater adaptability through supplementary documents (DO-330, DO-331, DO-332, and DO-333). The standard is not a prescriptive process model, but a process-oriented assurance framework that defines objectives to be satisfied, activities to be performed, and life cycle data to be generated and assessed for compliance. The extent and rigor of these depend on the Software Level—A to E—determined by the impact of software failure conditions on aircraft safety (from catastrophic to no effect).

The DO-178C standard establishes confidence in the correctness and reliability of a software component in proportion to the severity of its associated failure condition category, commonly referred to as its criticality level. To achieve this proportional assurance, DO-178C defines five distinct Design Assurance Levels (DALs). This classification framework serves as the foundation for assigning a DAL to a software component based on its role in, or contribution to, potential system-level failure conditions [1], [65]. Each DAL corresponds to the severity of the failure condition that may result from anomalous behavior in the software. These levels are described as follows:

- **Level A:** Associated with catastrophic failure conditions, which may lead to multiple fatalities and complete loss of aircraft. Components at this level are typically involved in safety-critical functions, such as flight control computers.
- **Level B:** Corresponds to hazardous or severe-major failure conditions. These reduce the flight crew's ability to operate the aircraft under adverse conditions and may result in serious or fatal injuries to a limited number of passengers or crew. Such failures can significantly degrade system performance, increase crew workload to unsustainable levels, and potentially lead to loss of control.
- **Level C:** Denotes major failure conditions that impair the crew's capability to manage the aircraft under adverse circumstances. This level includes failures that cause a notable reduction in functionality, increased crew workload, or physical distress to passengers and crew, but without posing a direct threat to aircraft integrity or survivability.
- **Level D:** Refers to minor failure conditions, which do not significantly affect the safety of the aircraft or its occupants. These conditions may lead to slight reductions in operational capability or comfort and may marginally increase the crew's workload. However, such events are considered manageable by the flight crew without affecting the overall safety of the flight.
- **Level E:** Represents software that has no effect on operational safety or aircraft functionality. Although software at this level does not impact safety margins, developers are still required to provide justification for this classification to the relevant certification authority, such as the Federal Aviation Administration (FAA).

DO-178C aligns the DAL levels with a set of process objectives that must be satisfied to demonstrate compliance and support certification. These objectives cover a wide range of development and verification activities, including requirements capture, design, coding, testing, configuration management, and quality assurance. The scope and rigor of these objectives increase proportionally with system criticality, ranging from 71 objectives for the highest assurance Level A to 69, 62, and 26 for Levels B, C, and D respectively, reflecting escalating safety assurance needs across DAL hierarchy.

DO-178C mandates a rigorous and structured verification framework for software assigned to Design Assurance Levels (DAL) A through D, with the depth and breadth of activities proportionate to the potential consequences of software failure. The standard requires a comprehensive suite of verification activities, including requirements-based testing, structural coverage analysis, and bidirectional traceability across all stages of the software life cycle. These verification tasks are designed to provide objective evidence that the software meets its requirements and performs reliably under all foreseeable conditions. The level of structural coverage required depends on the assigned DAL: Level A necessitates Modified Condition/Decision Coverage (MC/DC), Level B requires Decision Coverage, Level C demands Statement Coverage, and Level D has reduced verification obligations, limited to basic reviews and testing without structural coverage analysis.

Level E is distinct in that it is exempt from the verification objectives specified by the standard, as software at this level is determined to have no effect on the operational capabilities or safety of the aircraft. Nevertheless, depending on internal engineering practices, customer expectations, or organizational policies, certain development activities may still be voluntarily applied to Level E software to maintain uniformity across projects or to support future reclassification. Despite such discretionary efforts, the regulatory scrutiny for Level E software remains minimal compared to the higher assurance levels, reflecting its negligible impact on system safety.

### 3.1 Software Planning Process

DO-178C organizes its objectives, activities, and required data into five core software processes, ensuring that development and verification efforts are systematically planned, executed, and assessed. The first of these is the Software Planning Process, which establishes the foundation for all subsequent software activities. This process defines the overall approach to software development and verification, including the selection of life cycle models, identification of assurance levels, definition of applicable standards, and determination of tools and methodologies to be used throughout the development life cycle. It sets the scope and criteria for the certification effort and ensures alignment between stakeholders, developers, and certification authorities. To support this process, DO-178C requires the generation of specific planning documents, which form the basis for demonstrating compliance with the standard:

- The Plan for Software Aspects of Certification (PSAC) outlines the overall strategy for achieving certification, including the project's scope, DAL assignments, and the intended approach for satisfying DO-178C objectives.
- The Software Development Plan (SDP) details the methods, standards, and tools that will govern software requirements, design, implementation, and integration.
- The Software Verification Plan (SVP) specifies the verification activities, test environments, coverage objectives, and criteria for test success.
- The Software Configuration Management Plan (SCMP) defines how configuration items will be identified, controlled, and archived, ensuring the integrity of software baselines throughout the development life cycle.
- The Software Quality Assurance Plan (SQAP) describes the processes for monitoring compliance with approved plans and ensuring that quality objectives are consistently met.

These planning documents are not only critical for internal development discipline but also serve as a key interface with certification authorities, providing the necessary visibility and traceability into how DO-178C objectives will be achieved throughout the project. The planning process also includes the definition of software life cycle to be used (e.g., Waterfall, V-model, Agile) —which explains how the objectives of each software life cycle process will be satisfied.

### 3.2 Software Development Process

The Development Process under DO-178C encompasses the complete transformation of system-level requirements into a verified, certifiable software product. This process is structured into a series of interrelated activities—requirements capture, design, coding, and integration—each governed by rigorous standards to ensure traceability, correctness, and compliance with the assigned software level.

It begins with the capture and validation of High-Level Requirements (HLRs), which are derived from system specifications and define the intended software functionality without prescribing the implementation approach. These are then decomposed into Low-Level Requirements (LLRs), which describe the detailed, implementable logic necessary to realize the HLRs. Both HLRs and LLRs must be unambiguous, verifiable, and consistent, with complete bidirectional traceability maintained between them and the originating system requirements, as well as forward traceability to design, code, and verification artifacts.

Following requirements definition, the process advances to the development of the software architecture and detailed design. The architecture describes the structural organization of the software, including partitioning, control flow, data flow, and interfaces, ensuring compliance with safety and performance constraints. The detailed design phase specifies the algorithms, data structures, and inter-module interactions that will be implemented in source code, ensuring alignment with the allocated LLRs.

The coding phase implements design into source code in accordance with established standards, which address issues such as language subset usage, resource management, and avoidance of unsafe constructs. Each segment of code must be traceable to its corresponding LLR, and compliance with coding guidelines such as MISRA C++ [66] must be verified.

Subsequently, the integration phase combines individual software units into progressively larger assemblies until the complete executable software is formed. Integration ensures that all interfaces operate correctly, that functional behavior aligns with requirements, and that interactions between components do not introduce unintended behaviors.

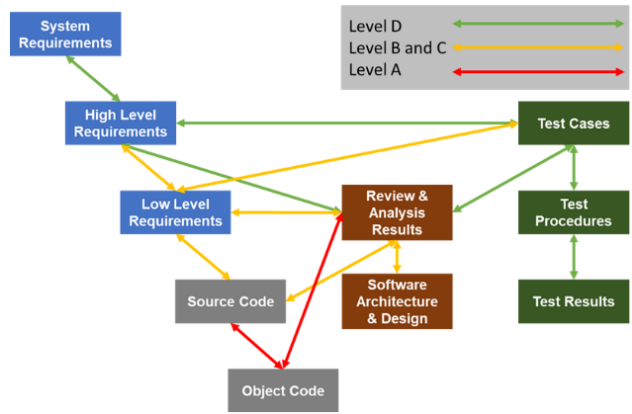

Figure 3: DO-178C traceability requirements as per the Design Assurance Levels.

Throughout the development process, DO-178C requires the creation and maintenance of specific data items that form part of the certification evidence. These include:

- Software Requirements Data defines High-Level Requirements (HLRs), which is the allocation of system requirements to software, including the derived safety-related requirements.
- Design Description specifies the software architecture and the Low-Level Requirements (LLRs) that will satisfy the High-Level Requirements.
- Software Standards defines the methods, rules and tools employed to develop the High-Level Requirements, software architecture, Low-Level Requirements, and the source code.
- Source Code is the human-readable implementation of the software.
- Executable Object Code is the compiled and linked binary ready for target execution.
- Trace Data includes the complete bi-directional traceability information linking High-Level Requirements (HLRs) to corresponding executable object code, in accordance with Design Assurance Level (DAL) assigned. The traceability requirements applicable for the specified DAL are illustrated in Figure 3.

All associated artifacts shall be maintained under configuration control in accordance with the control category prescribed for the assigned DAL. Furthermore, they shall exhibit complete traceability to demonstrate that every requirement has been correctly implemented, rigorously verified, and formally certified in compliance with the assurance level requirements defined in DO-178C. A visual representation of the decomposition process, showing how a system-level requirement is progressively refined into Low-Level Requirements (LLRs), is presented in Figure 4, using the 'Altitude Hold' function of an aircraft Flight Control System (FCS) as an illustrative example.

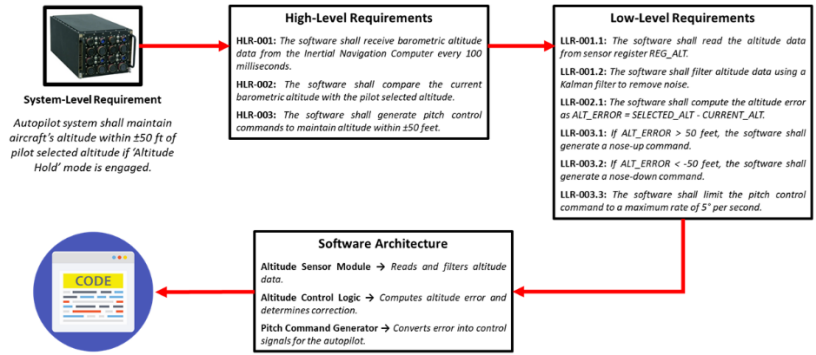

Figure 4: Software development process for aircraft autopilot altitude hold function.

### 3.3 Software Verification Process

The Verification Process in DO-178C is a structured and comprehensive set of activities aimed at ensuring that the developed software meets its requirements with the level of assurance appropriate to its criticality. This process operates in parallel with the development lifecycle, providing continuous evaluation and feedback to detect, isolate, and resolve defects as early as possible. It encompasses both reviews and analyses of development artifacts and test activities performed on integrated software, with the ultimate objective of demonstrating compliance with all applicable DO-178C objectives.

Verification begins with the review and analysis of requirements, design, and code. High-Level Requirements (HLRs) and Low-Level Requirements (LLRs) undergo rigorous evaluations to ensure they are accurate, complete, consistent, and verifiable, and that they maintain correct traceability to system requirements and to each other. Software architecture and detailed design are analyzed to confirm they are consistent with requirements, adhere to safety constraints, and avoid unintended functionality. Source code is reviewed to ensure compliance with coding standards, alignment with detailed design, and absence of constructs or practices that could lead to unsafe behavior, resource leakage, or timing anomalies.

Following the static verification phases, the process transitions to dynamic verification through testing. This includes unit testing of software components to confirm that each LLR has been correctly implemented, integration testing to ensure correct interaction between software units, and hardware / software integration testing to validate correct execution within the target environment. Finally, system-level testing is conducted to demonstrate that the complete software product satisfies the HLRs and performs as intended under operational conditions. For higher assurance levels (Levels A and B), DO-178C also mandates structural coverage analysis—including Statement, Decision, and Modified Condition / Decision Coverage (MC/DC)—to verify that all executable code has been exercised by tests and that no unintended, unverified code remains. Additionally, the standard mandates verification independence, whereby verification activities—such as reviews, analyses, and testing—are performed by personnel who are not the authors of the software being verified, thereby reducing the risk of bias and increasing assurance of correctness.

The Verification Process also ensures traceability between each requirement, its corresponding design and code implementation, and verification results that demonstrate its correctness. This bi-directional traceability is critical for ensuring that all requirements are implemented and tested, and all implemented functionality is justified by a requirement.

To facilitate software certification, DO-178C mandates the generation, review, and configuration-controlled maintenance of specific verification artifacts, including:

- Software Verification Results contains the documentation of the actual outcomes of verification activities, providing evidence that code coverage objectives have been achieved and that compliance with the applicable DO-178C objectives has been demonstrated.
- Software Verification Cases and Procedures includes the detailed specifications of the test inputs, execution steps, and expected outcomes for each verification activity.
- Trace Data includes the comprehensive bi-directional traceability linking each requirement to its corresponding design representation, source code implementation, and verification results, thereby demonstrating correctness and fulfillment of the intended functionality.

By enforcing both static and dynamic verification activities and by requiring rigorous documentation and traceability, the DO-178C Verification Process ensures that the delivered software not only fulfills its intended functionality but also exhibits the safety, reliability, and integrity demanded by airborne systems certification.

### 3.4 Software Configuration Management Process

The Configuration Management (CM) process is a critical assurance activity intended to preserve the integrity, consistency, and reproducibility of all software items throughout the software development lifecycle. The CM process provides a controlled framework to ensure that all configuration items are clearly identified, properly maintained, and systematically tracked, thereby enabling accurate reconstruction of any certified software state at any point in the lifecycle.

Key activities under this process include the identification of all software configuration items (SCIs), the establishment of version control mechanisms, and the implementation of formal change control procedures to evaluate, authorize, and record any modifications. Baselines are formally established and archived at defined lifecycle stages, with all subsequent changes meticulously documented to maintain traceability and prevent unauthorized alterations.

The required configuration management data, as prescribed by DO-178C, includes the following:

- Software Configuration Index (SCI) enumerates and describes all configuration items necessary to build software.
- Software Configuration Management Records provides a documented history of all approved modifications, including rationale and impact assessments.
- Problem Reports are a means to identify and record the resolution to software product anomalous behavior, process non-compliance with software plans and standards, and list any deficiencies in software life cycle data.
- Software Life Cycle Environment Configuration Index (SECI) identifies the configuration of the software life cycle environment, which is written to aid reproduction of the hardware and software life cycle environment for software regeneration, reverification, or software modification.

This disciplined CM approach ensures that software delivered for airborne systems meets the applicable assurance level requirements and remains under rigorous configuration control from inception through certification and maintenance.

### 3.5 Software Quality Assurance Process

Under DO-178C, the Software Quality Assurance (SQA) process serves as an independent assurance mechanism to provide objective evidence that all software development and verification activities are performed in accordance with approved plans, established standards, and the objectives defined in the standard. The SQA process is independent from development team, ensuring that evaluations are unbiased and that deviations or non-compliances are detected and addressed promptly.

The primary activities within this process include conducting formal audits of both development and verification tasks at planned intervals, reviewing project documentation for adherence to the approved Software Development Plan, Verification Plan, and other applicable lifecycle plans, and ensuring that all processes are executed in strict compliance with documented procedures. These audits verify that required DO-178C objectives are being met at the appropriate assurance level, and that anomalies are formally recorded, tracked, and resolved in a timely manner.

The required outputs of the SQA process, as specified by DO-178C, includes the Software Quality Assurance Records, which may consist of audit reports, completed checklists, non-compliance reports, and formal compliance assessments. These records serve as objective evidence for the certification authority that the software lifecycle processes have been followed as planned, that corrective actions have been implemented where necessary, and that the overall software product and its supporting processes meet the defined safety, quality, and assurance requirements.

### 3.6 Certification Liaison Process

The Certification Liaison process, as outlined in DO-178C, facilitates effective communication and coordination between the software development organization and the certification authority throughout the software lifecycle. Its purpose is to ensure that certification objectives are fully understood, appropriately addressed, and consistently demonstrated, thereby reducing the risk of misunderstandings or late-stage compliance issues. This process involves establishing and maintaining a structured communication channel to provide visibility into development progress, verification results, and compliance evidence at defined milestones.

Key activities in the Certification Liaison process include organizing and participating in formal reviews such as Stage of Involvement (SOI) audits, preparing and submitting the required certification data for review, and responding to certification authority queries and findings in a timely and traceable manner. The process also involves the coordination of problem reports, resolution tracking, and the provision of additional supporting evidence when required.

The principal deliverable of the Certification Liaison Process is the Software Accomplishment Summary (SAS), which serves as the primary artifact for demonstrating compliance with the Plan for Software Aspects of Certification (PSAC). This document, along with associated supporting records, constitutes objective evidence that the applicant has maintained transparent, structured, and traceable communication with the certification authority throughout the software development lifecycle. Furthermore, it confirms that all applicable DO-178C objectives have been comprehensively satisfied prior to the granting of certification. The integration of these activities with Stage of Involvement (SOI) audits ensures that each lifecycle phase is subject to formal review, thereby providing regulatory authorities with verifiable assurance of process compliance and product integrity.

### 3.7 Use of Tools and Supplements

DO-178C recognizes that software tools play a significant role in both development and verification activities, and it classifies them into two primary categories. Development tools are those capable of introducing errors into the airborne software, such as code generators or compilers, and therefore pose a potential risk to software integrity. Verification tools, on the other hand, do not introduce errors but may fail to detect them if they malfunction, as in the case of automated test frameworks or static analyzers. When organizations intend to rely on tool outputs without performing full manual verification, the tools must undergo Tool Qualification in accordance with the guidance provided in RTCA DO-330, ensuring that their operation is consistent, reliable, and appropriate for their intended purpose.

In addition to the core DO-178C guidance, several supplementary documents provide extended objectives and processes tailored to specific software development paradigms. DO-331 addresses Model-Based Development and Verification, offering additional guidance on the modeling process, simulation, and automatic code generation. DO-332 focuses on Object-Oriented Technology and Related Techniques, detailing how inheritance, polymorphism, and dynamic binding should be managed to maintain determinism and safety. DO-333 covers the application of Formal Methods, introducing mathematically rigorous approaches to software verification. Each supplement defines additional activities, objectives, and data requirements, enabling organizations to adopt these advanced methodologies while maintaining full compliance with DO-178C's safety and certification framework.

In summary, DO-178C provides a comprehensive assurance framework to ensure the safety and reliability of airborne software systems. For development agencies, compliance entails the rigorous execution of defined activities across planning, development, verification, configuration management, and quality assurance processes. Each process generates specific life cycle data that must be traceable, reviewable, and auditable. Emphasis on traceability, robust verification including structural coverage, and configuration integrity are central to achieving certifiability. While the standard allows flexibility in implementation, it demands stringent evidence for every objective claimed, forming the foundation for regulatory approval and operational deployment of safety-critical airborne systems. By enforcing this scalable assurance model, DO-178C ensures that software assurance efforts are commensurate with the potential safety consequences of failure, thereby supporting the development of safe and certifiable airborne software systems.

## 4 OVERVIEW OF AEROSPACE COMPANY AND SOFTWARE DEVELOPMENT PROCESSES

This section provides an overview of the aerospace software development organization under study, including its team structure and the pre-existing software development processes that were subsequently replaced by the Scrum framework. The organization is tasked with the development of software for Flight Control Computer (FCC), a critical component of the safety-critical Flight Control System (FCS) of an aircraft. The FCC software in question has been assigned Design Assurance Level A in accordance with DO-178C, as any malfunction or failure could lead to catastrophic consequences.

The company possesses specialized expertise in avionics flight control systems, having initially undertaken the full development of both the software and its associated hardware. Following the initial development and operational deployment, the organization assumes long-term responsibility for the software product, which includes performing corrective maintenance through bug fixes as well as adaptive and perfective maintenance in the form of feature enhancements. Over its two decades of operation, the company has accumulated extensive experience in the development and sustainment of FCC software across three distinct aircraft variants.

Historically, the organization has followed an annual release cycle, delivering a new software version approximately once every year. However, this extended release interval has occasionally been perceived as suboptimal by end-users and other stakeholders, as critical features are introduced too late to meet evolving operational needs. In certain cases, the intended functionality has lost its relevance entirely by the time of delivery due to rapid technological advancements, shifting market demands, and changing regulatory or mission requirements. This misalignment between delivery cadence and stakeholder expectations has been one of the driving factors behind the transition to agile development methodology.

### 4.1 Overall Team Structure

The software development organization within the company is deliberately structured around a diverse set of specialized roles, each of which plays a critical function in ensuring the successful delivery of safety-critical avionics software. In contrast to generic software teams, this structure reflects the stringent rigor, traceability, and compliance requirements dictated by DO-178C. The team composition typically encompasses Project Managers responsible for oversight and coordination, supported by five to ten System Engineers tasked with requirements engineering, safety analysis, and system-level validation. The core Software Development Team, generally comprising five to ten developers, is responsible for implementing avionics functions in compliance with coding standards and safety objectives. To enable effective verification, an additional five to ten engineers are dedicated to building and maintaining simulation and test environments essential for system-level evaluation. Furthermore, the organizational design preserves independence of assurance activities, in line with DO-178C, through specialized groups: three to five Configuration Management experts ensure baseline integrity and change control; three to five Software Quality Assurance (SQA) experts provide independent oversight of process adherence and compliance; three to five Software Analysis and Testing experts conduct verification activities, including static analysis, unit testing, integration testing, and robustness analysis; and three to five Certification Liaison specialists prepare certification data items, coordinate audits, and maintain structured communication with regulatory authorities. This multidisciplinary composition, balancing both development and independent assurance functions, not only satisfies the separation of responsibilities required under DO-178C but also exemplifies the high degree of rigor, collaboration, and systematic coordination inherent to aerospace software development.

Project managers hold overall responsibility for planning, execution, and control of the project. Their primary concern is ensuring that development activities are completed within the agreed timelines, allocated resources, and budgetary constraints, while delivering a software release accompanied by all required documentation in compliance with certification standards. System engineers function as critical interfaces between multiple stakeholders, including end users such as pilots, aircraft-level system engineers, hardware engineers, and the internal software development team. Their role encompasses requirements elicitation, clarification, and specification to ensure precision and completeness. They also provide technical interpretation of requirements to developers when necessary and are further responsible for conducting acceptance testing, thereby validating that the software conforms to the defined requirements. In this sense, system engineers serve as the principal communication channels with end users and customers.

Within the organizational framework, software developers form the core technical workforce, directly responsible for transforming validated system requirements into certifiable, executable avionics software. Their responsibilities extend beyond programming and include adherence to coding standards, participation in structured peer reviews, and preparation of the detailed technical documentation mandated under DO-178C, such as design descriptions, low-level requirements traceability, and verification records. Complementing their role, Software Quality Assurance (SQA) experts provide an independent oversight function, ensuring that all phases of the software life cycle strictly conform to DO-178C objectives. They achieve this through systematic process audits, compliance reviews, and assessments, thereby guaranteeing that regulatory and organizational standards are met throughout development. In parallel, Software Analysis and Testing experts fulfill the essential requirement for independent verification specified under DO-178C. Their activities include conducting rigorous test planning, unit and integration testing, static and dynamic analysis, robustness testing, and the preparation of verification evidence and test documentation that form part of the certification data set. Collectively, these roles establish a structured balance between development, oversight, and independent verification, a principle central to safety-critical assurances required by aerospace software certification standards.

In parallel, simulation and test environment engineers design and implement the sophisticated software and hardware environments required for integration and verification. Given the inherent complexity of aerospace systems, the Flight Control Computer (FCC) must interface not only with its immediate subsystems, such as sensors and control surfaces, but also with a wide range of other avionics components, including the Mission Management System (MMS), Inertial Navigation System (INS), and Electromechanical Management System (EMMS). Consequently, integration and system-level verification demand the development of complex, high-fidelity simulation and test environments that can accurately replicate these interactions. This parallel development effort is critical for enabling thorough validation of FCC functionality under representative operational conditions. Configuration Management (CM) experts are dedicated to ensuring compliance with configuration management processes, which include the identification of configuration items, version control, release management, and archival of baselines. The company places particular emphasis on configuration management, recognizing it as a cornerstone for maintaining integrity, traceability, and reproducibility of safety-critical software artifacts.

Finally, Certification Liaison experts serve as the coordinating body between the development team, system engineers, and certification authorities. They are responsible for ensuring that all required documentation and objective evidence is generated, maintained, and made available in accordance with certification requirements. In addition, they facilitate communication with certification agencies and support Stage of Involvement (SOI) audits and reviews, thereby providing assurance that all DO-178C objectives are being consistently addressed. This carefully structured distribution of roles illustrates how the organization integrates technical expertise, regulatory compliance, and systematic project management to deliver safety-critical avionics software. The overall working model of the company, including the interaction of various roles and processes, is depicted in Figure 5.

### 4.2 Legacy Software Development Process

The company adheres to the V-model development approach (Figure 2), which represents a structured and rigorous extension of the traditional Waterfall model, tailored for safety-critical aerospace systems. This model emphasizes systematic progression from requirements definition to implementation, with corresponding verification and validation activities aligned at each stage. The development lifecycle begins with the reception of formal, executive-level requirements from the customer organization. Upon project initiation, a project manager is formally assigned to lead the effort. With support from the multidisciplinary technical team, a comprehensive project plan is developed, reviewed, and mutually agreed upon with customer executives to ensure alignment of expectations regarding scope, milestones, resource allocation, and deliverables. Once initiated, the System Engineering team engages extensively with aircraft-level system engineers, pilots, and other relevant customer stakeholders to elicit, refine, and specify the High-Level Software Requirements (HLRs). This stage is characterized by rigor and intensity, sometimes spanning up to six months due to the safety-critical nature and complexity of avionics systems. Requirements may be specified in natural language, structured formats, or formal modeling notations such as UML statecharts, activity diagrams, or sequence diagrams to capture system behavior with precision. Following iterative reviews and deliberations involving system engineers, software developers, and other functional groups, HLRs are baselined, thereby precluding further modifications unless formally justified through a controlled change management process. This freeze is necessary to maintain stability in downstream activities.

Following the definition and validation of High-Level Requirements (HLRs), the software development team undertakes their systematic decomposition into Low-Level Requirements (LLRs) and subsequently derives a detailed software design. These artifacts constitute the formal foundation for subsequent phases, including software implementation, unit testing, and code-level verification. In parallel, the simulation and test environment development team derive the associated test requirements and construct the simulation infrastructure and monitoring systems necessary to enable integration testing and system-level validation of the avionics software. Complementing these efforts, the software analysis and testing team develops the verification plan, including detailed test plans, test cases, and automated test scripts, in alignment with DO-178C objectives for independent verification. Given the inherent complexity of safety-critical flight control systems, the simulation environment frequently incorporates hardware-in-the-loop (HIL) configurations, which enable a realistic replication of operational scenarios and thereby provide a high degree of assurance regarding system behavior under real-world conditions. Once software implementation is finalized—supported by unit testing, static code analysis, and structured peer reviews—the software build is integrated into the simulation and test environment and subsequently deployed onto the target Flight Control Computer (FCC) hardware. At this integration stage, system engineers assume responsibility for conducting acceptance testing and validation, ensuring that the implemented software not only satisfies the established HLRs but also aligns with operational and end-user expectations. Notably, in legacy framework, the final acceptance test plan is authored and maintained by the system engineering team, rather than the software analysis and testing group, thereby reflecting the system-level ownership and accountability required for flight-critical avionics software development under DO-178C.

Following internal validation, aircraft-level system engineers and pilot representatives are engaged to conduct operational validation, providing domain-specific feedback based on actual usage contexts. If any deviations, shortcomings, or misalignments with user requirements are identified, a rework cycle is initiated, requiring partial or complete iteration of the design and development phases. This rework is followed by comprehensive regression testing to ensure that corrective changes do not adversely impact existing functionality. The software is finally delivered to aircraft-level system engineers, who conduct comprehensive hardware-in-the-loop integration testing across avionics subsystems at another facility. This phase is followed by flight testing, where the software is evaluated under actual operational conditions to confirm performance, robustness, and safety. Meanwhile, the Configuration Management (CM) team establishes and enforces the configuration management plan, encompassing identification of configuration items, version control, baselining, release management, and archival. In parallel, Software Quality Assurance (SQA) experts provide an independent oversight function, verifying that all life cycle processes—ranging from requirements capture and design, through implementation and verification, to configuration management—are executed in compliance with the defined plans, standards, and procedures. Their audits, reviews, and process assessments ensure that every artifact remains fully documented, objectively evaluated, and traceable to the applicable certification objectives, thereby upholding the independence requirements mandated by DO-178C. The organization's confidence in the V-model stems from over two decades of experience applying this methodology to flight-critical systems. Continuous training, accumulated expertise, and lessons learned across multiple aircraft programs have reinforced the team's capability to apply the V-model effectively in compliance with the stringent objectives of DO-178C.

### 4.3 The Certification Process and Regulatory Oversight

The certification of aircraft software is governed by Federal Aviation Administration (FAA) Order No. 8110.49 [67], which mandates adherence to industry standards such as RTCA DO-178C (or ED-12C in Europe) under the oversight of regulatory bodies like the FAA. Central to this process are four mandatory audit points, known as Stages of Involvement (SOI), during which developers present evidence of compliance to certification authorities for independent review. Because DO-178C prescribes a process-oriented framework, these reviews must be meaningfully integrated throughout the software life cycle to establish confidence in both the processes employed and the resultant product. Accordingly, maintaining regular contact between the applicant and the FAA is critical to ensuring transparency, confidence, and early detection of potential compliance issues.

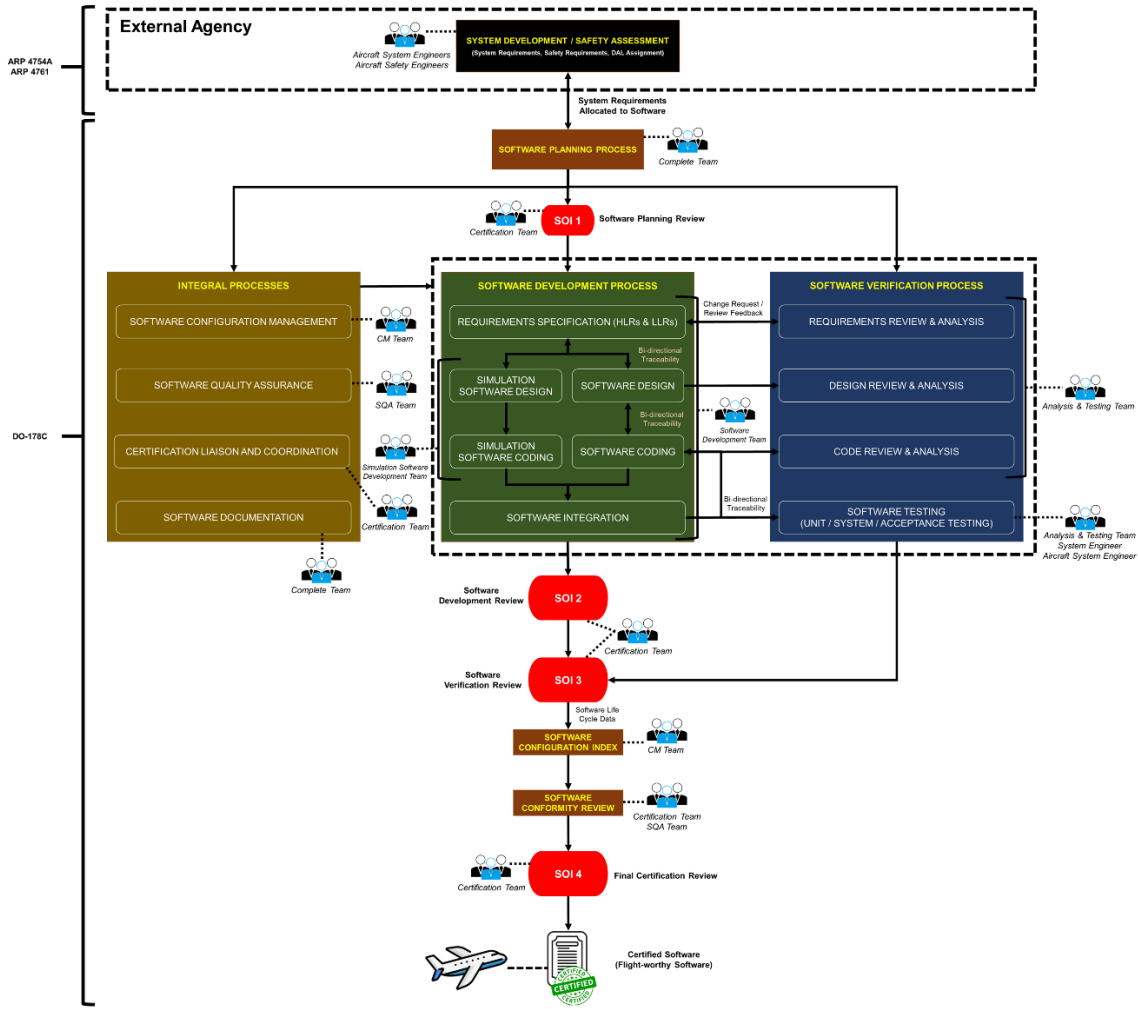

Figure 5: Legacy aerospace software development and certification processes aligned with DO-178C objectives and FAA SOI reviews.

The four primary review stages are defined as follows:

1. **Software Planning Review (SOI #1):** Conducted upon completion and review of all the planning documents.
2. **Software Development Review (SOI #2):** Conducted once a representative portion (typically ≥ 50%) of software development artifacts—such as requirements, design, and code—have been produced and reviewed.
3. **Software Verification Review (SOI #3):** Conducted when at least half of the verification and testing artifacts are complete and reviewed.
4. **Final Certification Review (SOI #4):** Conducted after final build is completed, verification is finalized, conformity checks are performed, and the software application is ready for formal system-level certification approval.

While these four reviews form the standard framework, FAA Order No. 8110.49 [67] clarifies that the number and type of reviews may vary depending on project characteristics. Certain projects may consolidate reviews, whereas others—particularly those involving novel technologies or higher-risk software—may require additional oversight. In practice, if all DO-178C objectives are demonstrated to the satisfaction of the authorities across these reviews, the software is approved as part of the aircraft's Type Certification. The Certification Liaison team in aerospace company under study assumes responsibility for initiating coordination with the regulatory authority after project commencement, which includes establishing the certification basis and planning the SOI reviews in alignment with DO-178C objectives. Furthermore, the team maintains continuous coordination with the certification authority to support the effective conduct of these reviews. This includes overseeing the preparation, validation, and finalization of certification artifacts, which serve as the objective evidence required to demonstrate compliance with DO-178C. By serving as the interface between the development organization and the certification authority, the Certification Liaison team facilitates clarity in expectations, consistency in process adherence, and timely resolution of compliance-related issues throughout the software life cycle.

The extent and nature of authority involvement are determined by multiple factors, including software level (e.g., Level A safety-critical systems), product complexity, novelty of methods or models, prior certification experience of the applicant, and service history of related systems [67]. In the case under study, the organization develops Level A software and possesses extensive experience in certifying similar systems. Consequently, the certification authority typically assigns a medium level of involvement, consisting of three desk reviews (SOI #1, SOI #2, SOI #3) and one on-site review at project completion (SOI #4). This balance of desk and on-site reviews provides regulatory assurance while optimizing resource utilization for both parties. This structured approach to SOI reviews underscores how a balance between regulatory oversight and organizational maturity can streamline certification without compromising compliance rigor, thereby enhancing both efficiency and audit credibility in safety-critical aerospace software development. The complete process adopted by the company, including its alignment with certification audits (SOI reviews), is illustrated in Figure 5.

**4.4 Challenges of Traditional Software Development Process**

Despite the systematic rigor of the V-model, the company encounters inherent limitations associated with traditional Waterfall-derived processes. Several challenges were observed:

- Excessive focus on upfront planning and documentation – In some cases, the planning and documentation phases consumed more effort and time than actual software development. This phenomenon, often referred to as analysis paralysis, hindered agility and responsiveness to evolving needs.
- Fragile or incomplete initial requirements – Given the long duration of the development cycle, requirements defined at project inception often proved incomplete or outdated. Any ambiguities or oversights in requirements propagated downstream, resulting in costly redesigns and delays.
- Late integration and feedback – Since the model is non-iterative and non-incremental, the integrated product was only made available to stakeholders toward the end of the lifecycle. Consequently, critical feedback from pilots and operational experts was received very late, often during flight testing. By this stage, implementing design modifications or addressing defects required disproportionately high effort and incurred significant costs.
- Dependency-driven project delays – The delayed availability of deliverables from one subsystem caused cascading delays in other subsystems dependent on its outputs, further amplifying schedule risks across the program.
- Delayed delivery of new features and bug fixes – End-users, including pilots and program managers, expressed dissatisfaction with the long release cycles. New software releases—whether addressing defects from previous versions or introducing new features—typically reached production only once every two years. This extended cycle involved complete system integration with other avionics subsystems at a separate integration facility, followed by aircraft-level testing and validation before final deployment. By the time integration was completed, some requirements had lost relevance due to technological advancements or evolving operational needs.

Overall, while the V-model provided discipline, traceability, and compliance with DO-178C objectives, it also imposed rigidity, inefficiency, and delayed responsiveness to stakeholder needs. These shortcomings highlighted the necessity of exploring more adaptive, iterative approaches, such as Agile-based methods, which could deliver incremental value, accelerate feedback loops, and reduce rework costs in the context of aerospace software development.

## 5 TRANSITION TO AGILE AND ADOPTION OF TAILORED SCRUM FRAMEWORK

This section presents a comprehensive account of the structured steps undertaken by the company in transitioning from the traditional V-model to Agile practices. It details the practical implementation of the Scrum framework, carefully adapted to align with the stringent requirements of aerospace software development. The narrative highlights how organizational processes, roles, and workflows were redefined to ensure compliance with DO-178C objectives while simultaneously achieving the flexibility, incremental delivery, and rapid feedback cycles enabled by Agile. The description also emphasizes the mechanisms through which Scrum was tailored to address domain-specific constraints, including safety-critical certification needs, extensive verification activities, and coordination with system-level stakeholders.

### 5.1 Facilitating Agile Transformation Through Awareness and Training

As highlighted in Section 4, the organization's engineering and management teams had accumulated more than two decades of experience working within the traditional V-model framework. Consequently, transitioning to Agile necessitated a comprehensive change management effort to ensure alignment and buy-in across all stakeholders. In line with the observations specified by Islam and Storer [6] and Benedicenti et al. [96], which underscore the importance of addressing cultural resistance as a prerequisite for the effective adoption of Agile methodologies, the company undertook a series of structured awareness sessions. These included technical presentations and deliberative discussions with senior management, system engineers, software developers, and other key personnel. The purpose of these engagements was to articulate the limitations of the existing model, to demonstrate how Scrum could alleviate those challenges, and to address concerns raised by stakeholders through open dialogue. Following the resolution of these concerns, formal approval was obtained from executive leadership to proceed with the Agile transition. Subsequently, specialized training sessions were delivered by domain experts to build organizational competence in Scrum practices. This was particularly critical for system engineers, many of whom had limited prior exposure to Agile concepts, thereby ensuring that entire development team possessed a consistent and comprehensive understanding of the methodology before its formal deployment.

### 5.2 Adaptation of Scrum Roles in the Context of Aerospace Software Development Under DO-178C

The Scrum Guide [23] formally delineates three foundational roles that constitute the operational core of the Scrum framework: Product Owner, Scrum Master, and Development Team. As emphasized by Schwaber and Beedle [30], and further substantiated by Lei et al. [39], Scrum demonstrates its greatest effectiveness when applied within small, cross-functional, and self-organizing teams, typically consisting of three to nine members. Nevertheless, the distinctive characteristics of safety-critical aerospace software development, together with rigorous compliance objectives mandated by DO-178C, necessitated both a redefinition and an extension of these canonical Scrum roles. To ensure compatibility with regulated aerospace domain, the organization adopted a hybrid adaptation of Scrum in which standard roles were tailored to incorporate regulatory compliance responsibilities, while also preserving essential functional departments that traditionally operate within aerospace projects. Specifically, teams such as Configuration Management, Software Quality Assurance (SQA), Software Analysis and Testing, and Certification Liaison were retained in their established capacities, thereby minimizing organizational disruption and ensuring continuity of compliance-critical processes.

Preserving these independent roles was particularly crucial because DO-178C explicitly requires separation of responsibilities—for instance, independent verification, objective quality assurance, and structured certification coordination—which could not be fully embedded within the development team without compromising compliance integrity. This deliberate integration thus enabled the company to maintain alignment with DO-178C objectives, safeguard audit credibility, and uphold independence of verification, while still realizing the flexibility, responsiveness, and iterative benefits of the Scrum methodology. In addition to these preserved functions, a dedicated Documentation Team was established with the explicit mandate of preparing, consolidating, and maintaining all certification documentation required under DO-178C. This structural modification was introduced to enable developers to focus exclusively on working software, in accordance with Agile principles, without being overburdened by the extensive documentation demands of certification. In prior organizational practice, documentation responsibilities had been distributed across multiple teams, often diluting focus and introducing inefficiencies. By centralizing this responsibility, the company achieved both improved process clarity and greater alignment with Agile values, while ensuring that documentation remained comprehensive, consistent, and certification-ready.

With respect to the Scrum Master role, the organization initially encountered a significant challenge due to its limited internal expertise in Agile and Scrum methodologies. To address this gap, an adjunct Subject Matter Expert (SME) was engaged, providing specialized guidance, mentorship, and oversight during the formative stages of Scrum adoption. Furthermore, given the stringent demands of DO-178C compliance, it was deemed imperative that the individual fulfilling the Scrum Master role possess not only a deep understanding of Agile facilitation and process leadership but also a strong familiarity with DO-178C objectives, certification requirements, and regulatory interactions. This dual expertise was considered essential to ensure that both the agility of the development process and the rigor of certification activities were consistently upheld throughout the software life cycle. This external expertise was instrumental in ensuring that Scrum ceremonies, principles, and practices were not only correctly implemented but also appropriately adapted to the constraints of a safety-critical aerospace environment, thereby avoiding any conflict with regulatory obligations.

In Agile practice, the Product Owner (PO) is conventionally tasked with defining, prioritizing, and managing the Product Backlog Items (PBIs). Typically, this role is performed by a single individual who serves as the representative voice of multiple stakeholders. While this single-point ownership approach may function effectively in non-safety-critical domains, it is insufficient in the context of safety-critical aerospace projects governed by DO-178C, where the complexity and criticality of requirements demand broader expertise and multiple perspectives. Ribeiro et al. [5], in a post-mortem analysis of Waterfall-based aerospace projects, observed that POs frequently overlooked technical details when formulating PBIs, which in turn contributed to project delays—an inherent limitation of concentrating responsibility in a single role. To address this challenge, Sutherland et al. [72] emphasized the importance of exploring requirements collaboratively as a team. They further asserted that, to be considered ready, each PBI must satisfy established criteria before it can be selected as a candidate backlog item during Sprint Planning [72], [73]. Building on this principle, the traditional PO role was adapted into a multi-disciplinary Product Ownership Team, comprising company system engineer, an aircraft-level system engineer, and a representative end-user (pilot). This collaborative team assumed responsibility for the definition, refinement, and prioritization of the Product Backlog, with a particular focus on safety requirements. At this level, not only were the high-level requirements elicited and expressed in the form of user stories, but explicit consideration was given to functional, interface, safety, robustness, performance, and non-functional requirements. The integration of domain-specific knowledge from both engineering and operational perspectives enabled the Product Ownership Team to generate a sufficiently refined backlog ahead of Sprint Planning, thereby ensuring readiness for subsequent development, verification, and compliance activities. To formalize this process, a Definition of Ready (DoR) was established for PBIs, ensuring that only well-prepared backlog items were eligible for selection during Sprint Planning.

The Development Team was restructured to reflect both Agile collaboration principles and DO-178C independence requirements. The team composition included:

- Software development sub-team focusing on avionics software implementation.
- Simulation software development sub-team dedicated to simulation and test software development, enabling verification in both virtual and hardware-in-the-loop environments.
- Configuration management sub-team, ensuring version control, build integrity, the impact and compliance implications of new requirements before their integration into the next sprint, and compliance with DO-178C's configuration baselines.
- Analysis and testing sub-team, mandated by DO-178C to provide unbiased verification through static and dynamic code analysis, software testing, and coverage analysis.
- Certification sub-team, responsible for coordinating with certification authorities, verifying process conformance to DO-178C objectives, preparing certification evidence, and managing Stage of Involvement (SOI) audits.
- Documentation sub-team, tasked with producing and maintaining technical documentation, including requirements traceability matrices across all lifecycle artifacts, which could not be comprehensively handled by developers due to the holistic, system-wide perspective required.
- SQA sub-team ensuring that all the software life cycle activities and outputs comply with the approved plans, standards, and procedures.

Verification and compliance-related activities were distributed across roles to ensure independence of verification, as prescribed by DO-178C. For instance, static and dynamic code analysis, software testing, and coverage analysis were performed by the independent analysis and testing sub-team, while traceability documentation was prepared by the documentation sub-team. The certification sub-team coordinated with regulatory bodies for reviews and certification audits, while concurrently monitoring development activities to ensure continuous compliance with certification objectives. Through this structured redefinition of roles, the company achieved a balance between Scrum's collaborative and iterative strengths and the formal rigor of DO-178C compliance, thereby enabling Agile methods to be effectively adapted for safety-critical aerospace software development. The composition of these adapted roles—reflecting the integration of Agile collaboration principles with DO-178C compliance requirements—is illustrated in Figure 6, which depicts the expanded Scrum team structure encompassing multi-disciplinary product ownership, dual-skilled Scrum facilitation, and specialized sub-teams to ensure both agility and certification integrity.

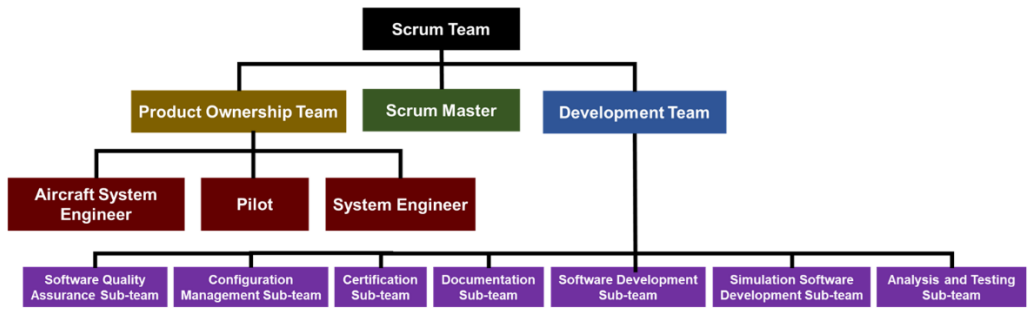

Figure 6: Adapted Scrum team structure for safety-critical aerospace software development under DO-178C.

## 5.3 Novel Adaptation of Scrum for Safety-Critical Aerospace Software

For the initial project implementation, the team elected to employ Scrum Type A, unlike the Scrum4DO178C process proposed by Ribeiro et al. [60] which modified Scrum Type C, primarily because the team was relatively new to Scrum and required a structured, clearly bounded approach. Type A emphasizes strict separation between Sprints, with all planned work confined to the Sprint timebox and minimal overlap across iterations [74]. This model provided the necessary clarity and predictability for a team in its formative stages of Agile adoption, particularly in the aerospace domain where compliance, documentation, and process discipline are critical. The controlled cadence of Type A facilitated learning, alignment with certification requirements, and reduced the risk of scope mismanagement. However, as the team's familiarity with Scrum practices matured and their capability for parallel workstreams improved, the rigid boundaries of Type A were found to constrain productivity and delivery speed. Consequently, the team transitioned to Scrum Type C, which embraces extensive overlap between phases of work and enables near-continuous development and integration. Type C—while more demanding in terms of automation, backlog management, and team discipline—offered the efficiency and responsiveness needed to align with the evolving complexity of aerospace software development. This shift reflects a natural progression in Scrum adoption, where organizations begin with a more sequential model and advance towards highly concurrent, continuous-delivery practices as their Agile maturity increases.

The project commenced with the Product Ownership (PO) Team assuming primary responsibility for defining, structuring, and prioritizing the Product Backlog, which, in this adaptation, was aligned to serve as the Agile equivalent of the High-Level Requirements (HLRs) prescribed by DO-178C. The PO Team received system-level requirements from the aircraft-level system engineers, typically supported by key artifacts such as the Pilot Operating Procedures (POP), which specify pilot actions in the cockpit and corresponding system responses; System Design Descriptions (SDD), which describe subsystem communication and operational logic; and Interface Control Documents (ICD), which define data exchanges between avionics subsystems through protocols such as MIL-STD-1553 and RS-422. Each backlog item was meticulously analyzed and assigned story points, which provided a relative measure of both implementation effort and inherent complexity. The assignment of story points was not arbitrary; rather, it was systematically grounded in the development team's collective historical experience with comparable tasks, thereby ensuring a consistent and realistic basis for workload estimation. Crucially, the allocation of story points encompassed the entire lifecycle of activities required for the successful realization of a user story. This included software architecture and design, coding, verification and validation activities, test case development, creation of test environments, traceable documentation, configuration management, and software quality assurance (SQA). To ensure rigor and transparency, relative estimation techniques were employed to account for complexity, risk, dependencies, and implementation effort. Furthermore, consensus-based methods such as Planning Poker [68], [69] were applied, enabling the development team to evaluate user stories against a base reference, discuss differing perspectives, and converge on a shared estimation outcome. This structured approach minimized bias, promoted team-wide alignment, and enhanced estimation accuracy.

The Product Ownership Team documented requirements that were unambiguous, complete, and well-understood as user stories, while deferring those with uncertainties or incomplete specifications to later iterations. This deferred elaboration allowed requirements to be progressively refined and clarified over the course of the project, ensuring accuracy without impeding early progress. The estimation and prioritization process were conducted collaboratively by the Product Ownership Team—comprising system engineers, a pilot, and an aircraft-level system engineer—together with the full development team. This multidisciplinary participation was critical to ensuring that both technical expertise and operational perspectives were adequately represented. Prioritization of backlog items was guided by several decisive factors:

- Functional dependencies between items, ensuring enabling functionality was developed prior to dependent tasks.
- Operational importance from the pilot's perspective (e.g. prioritizing features that directly influence cockpit usability and flight operations).
- System integration considerations identified by the aircraft-level system engineer (e.g. prioritizing functionalities whose early implementation enabled integration and verification with other avionics systems such as multifunction displays and mission management computer).

As an illustration, within the autopilot functionality of the Flight Control Computer (FCC), the pilot might prioritize the Altitude Hold mode due to its direct operational significance, whereas the aircraft-level system engineer might emphasize early development of cockpit interface processes to enable end-to-end validation of autopilot initiation across multiple avionics subsystems. In this manner, backlog prioritization balanced operational relevance with integration readiness, ensuring that both end-users need and system-level dependencies were addressed in tandem. Finally, once user stories (representing HLRs) were defined, they underwent rigorous review processes to ensure completeness, clarity, consistency, verifiability, and conformance to applicable standards. This review cycle provided an additional assurance mechanism, aligning the Agile backlog with the stringent requirements of DO-178C compliance while retaining the adaptability and iterative refinement central to Scrum practices.

In parallel, the Certification sub-team initiated formal coordination with the regulatory authority. This entailed establishing the certification basis, defining the schedule for Stages of Involvement (SOI) reviews, and presenting the proposed process adaptations that reconciled Agile methods with DO-178C objectives. During these discussions, the certification authority was briefed on the adoption of Scrum practices, including incremental and parallel execution of development and verification activities. Given the introduction of this novel process model, the authority elevated its level of involvement from Medium to High [67], requiring four on-site reviews: Software Planning Review (SOI #1), Software Development Review (SOI #2), Software Verification Review (SOI #3), and the Final Certification Review (SOI #4). However, recognizing that Scrum does not adhere to a rigid, sequential waterfall model, the authority agreed that SOI #2 and SOI #3 would be conducted after the completion of approximately half of the planned sprints. This compromise accommodated the incremental nature of Agile delivery while preserving oversight at critical lifecycle phases.

Simultaneously, the system engineers and the development team jointly initiated the preparation of DO-178C-compliant planning artifacts, which provide the foundation for traceability, certification readiness, and process rigor throughout the project lifecycle. The documentation team assumed primary responsibility for this task, consolidating inputs from all relevant stakeholders to generate precise and auditable planning documents. The details of Agile adoption and Scrum implementation tailored to the aerospace domain were formally captured in the Software Development Plan (SDP). The planning phase included the development of the following key certification artifacts:

- **Plan for Software Aspects of Certification (PSAC):** Prepared by documentation sub-team based on inputs from certification sub-team, outlining the certification approach, objectives, and planned Stages of Involvement (SOI).
- **Software Development Plan (SDP):** Prepared by the documentation sub-team based on inputs from the software development sub-team, detailing processes, tools, and methods for requirement decomposition, design, coding, and integration within the Scrum-based workflow.
- **Software Verification Plan (SVP):** Prepared by the documentation sub-team based on inputs from the analysis and testing sub-team, in conjunction with the simulation software development sub-team, specifying verification strategies, test case derivation, and validation mechanisms.
- **Software Configuration Management Plan (SCMP):** Prepared by the documentation sub-team based on inputs from the configuration management sub-team, defining baselining, version control, release procedures, and archival mechanisms to ensure configuration integrity.
- **Software Quality Assurance Plan (SQAP):** Prepared by the documentation sub-team based on inputs from the software quality assurance sub-team, describing independent audits, reviews, and assessments to verify compliance with DO-178C objectives across all lifecycle processes.

The initial phase of Product Backlog Item (PBI) creation and overall planning spanned approximately one month. By the conclusion of this period, the activities associated with backlog definition in Scrum [23] had been completed in parallel with the Software Planning Process prescribed under DO-178C [1]. This dual accomplishment signified that both the Agile framework's preparatory stage and the regulatory compliance planning phase had reached maturity. Following the completion of these foundational activities, the Certification sub-team initiated preparations for the first Stage of Involvement (SOI) audit, which serves as a formal checkpoint for verifying planning compliance with certification objectives [1], [67]. This alignment between Agile planning milestones and certification requirements ensured that the project proceeded in a structured, compliant, and auditable manner from the outset.

To illustrate, at the system-level, one of the key requirements related to Autopilot Mode Functions was captured in Scrum as an Epic:

"As a pilot, I want the Flight Control Computer to support multiple autopilot modes (e.g., Altitude Hold, Heading Hold, Airspeed Hold, Approach Hold, and Navigation Hold), so that I can reduce workload and enhance flight safety."

This Epic was subsequently decomposed into user stories (HLRs) and organized into the Product Backlog, complete with detailed acceptance criteria. Notably, acceptance criteria were structured on two levels: (i) functional acceptance, performed by the Product Ownership team to validate operational correctness, and (ii) compliance acceptance, conducted by the Certification sub-team to ensure adherence to all applicable DO-178C objectives. This dual-layer acceptance mechanism guaranteed that every increment was not only operationally functional but also demonstrably compliant with certification standards. Moreover, by archiving acceptance evidence at each iteration, the process created an auditable trail, facilitating eventual certification while preserving the incremental value delivery central to Scrum.

To ensure that all certification objectives were systematically addressed throughout the development lifecycle, a dedicated set of High-Level Requirements (HLRs) was defined to represent certification and compliance activities within the Agile framework. These certification-related HLRs were expressed as user stories—such as those for traceability, verification evidence generation, configuration management, data pack compilation, and audit readiness—to embed DO-178C objectives directly into the Scrum workflow. By managing these certification-oriented requirements as Product

Backlog Items (PBIs), the team ensured that compliance tasks were planned, estimated, and completed within individual sprints alongside functional development. This approach enabled continuous certification readiness, supported evidence generation for Stage of Involvement (SOI) reviews, and ensured that all verification and documentation deliverables were available for final audit without deferring compliance activities to post-development phases.

The subsequent phase involved Sprint Planning, during which the development process was structured into iterations of four weeks in duration. The selection of a four-week sprint cycle was informed by the Scrum Guide [23], which recommends iterations of two to four weeks as an effective means to enhance team focus, maintain consistency, enable incremental value delivery, and facilitate regular inspection and adaptation by both the team and stakeholders. Adopting a four-week sprint duration, however, necessitated the ability to decompose requirements into appropriately sized increments that could be feasibly completed within the defined timeframe. This required the breakdown of higher-level requirements into smaller, more precise backlog items, each representing a scope of work aligned with the team's capacity and the sprint boundaries. The primary rationale for selecting a four-week sprint, as opposed to a shorter two-week iteration, was directly tied to the stringent compliance obligations mandated by DO-178C. Specifically, a sprint would only be deemed complete if not only the implementation but also the full spectrum of compliance activities—including reviews, traceability updates, and verification tasks—were completed for all items within the sprint backlog. This alignment between Scrum iterations and certification requirements ensured that each increment delivered was both functionally operational and certification-ready, thereby reducing compliance risks at later stages of the development lifecycle.

Within the Scrum framework, Sprint Planning is a collaborative activity involving three key roles: the Product Ownership Team, the Scrum Master, and the Development Team. The Product Owners were responsible for articulating the Sprint Goal and clarifying the scope of Product Backlog Items (PBIs). The Scrum Master facilitated the session, ensuring adherence to Scrum principles and smooth execution of the planning process. The Development Team then evaluated its available capacity and committed to the selected PBIs, determining how these would be delivered during the sprint. A critical outcome of the initial Sprint Planning was the formal establishment of the Definition of Done (DoD), which served as the agreed benchmark for completion. In this context, a user story was deemed "done" only when it satisfied the full spectrum of activities required for compliance with DO-178C. These included decomposition into low-level requirements, development of software architecture and design (e.g., state charts, class diagrams, flow charts, pseudocode, sequence diagrams, data flow diagrams, and control flow diagrams), coding and implementation in compliance with standards such as MISRA C/C++, execution of static analysis, unit testing and system-level testing, structural coverage analysis, preparation of all relevant documentation (design reports, compliance evidence, and traceability matrices from HLRs to object code), validation through integration testing in both simulated environments and with the actual hardware, as well as completion of configuration management, Software Quality Assurance (SQA), and verification of all applicable DO-178C objectives.

Building upon the established framework, the Product Ownership Team finalized the Sprint Backlog through close collaboration with all participating teams. The selection of Product Backlog Items (PBIs) was informed by three primary parameters: the team's empirically demonstrated velocity, their projected capacity for the sprint, and the availability of required human resources. This systematic and evidence-driven approach ensured that the commitments undertaken during sprint planning were both realistic and achievable within the designated timeframe. As an illustrative case, in one of the initial sprints, the Sprint Goal was defined as the design, implementation, documentation, and verification of the Altitude Hold mode within the Flight Control Computer (FCC). The selected backlog items spanned from HLR-1 (Altitude Hold Mode Engagement) to HLR-11 (Sensor Data Failure Handling) (refer to Appendix A for the complete list of user stories / HLRs). Given that this initiative marked the team's first attempt to apply Scrum practices within a safety-critical context, the scope of initial sprint was deliberately constrained. This conservative planning strategy allowed the team to gradually adapt to tailored Scrum process, establish a sustainable working rhythm, and build momentum for subsequent iterations, while simultaneously ensuring adherence to rigorous development and verification standards mandated by DO-178C.

Following the finalization of the Sprint Backlog, the selected PBIs were systematically decomposed into smaller, manageable tasks, each mapped to activities explicitly required under DO-178C. This decomposition served several critical purposes: it enhanced the granularity and accuracy of effort estimation, enabled balanced workload distribution among teams, and improved progress monitoring throughout the sprint. Furthermore, aligning task structures with DO-178C objectives ensured that all safety-critical development and verification activities were comprehensively addressed, thereby reinforcing compliance with certification requirements. For instance, the user story HLR-1—Altitude Hold Mode Engagement, corresponding to the implementation of the Altitude Hold function (refer Appendix A for details) was decomposed into a structured set of tasks, including but not limited to the following:

- Develop Low-Level Requirements – Software Development Sub-team
- Review and analyze Low-Level Requirements – Software Development Sub-team
- Create software architecture and design (flowcharts, UML diagrams) – Software Development Sub-team
- Review and analyze software architecture and design – Software Development Sub-team
- Implement / code LLR-1.1 Command Reception – Software Development Sub-team
- Implement / code LLR-1.2 Target Altitude Capture – Software Development Sub-team
- Implement / code LLR-1.3 Mode Enable Logic – Software Development Sub-team
- Implement / code LLR-1.4 Engage Interlock – Software Development Sub-team
- Implement / code LLR-1.5 Persistence and Baseline – Software Development Sub-team
- Review and analyze the source code – Software Development Sub-team
- Develop simulation software requirements – Simulation Software Development Sub-team
- Create simulation software design – Simulation Software Development Sub-team
- Implement / code simulation software – Simulation Software Development Sub-team
- Carry out static code analysis (MISRA C++ compliance and complexity checks) – Analysis and Testing Sub-team
- Develop and execute unit tests (100% MC/DC) – Analysis and Testing Sub-team
- Develop and execute integration tests (100% MC/DC) – Analysis and Testing Sub-team
- Develop and execute system tests (100% MC/DC) – Analysis and Testing Sub-team
- Review and analyze verification activities, including test cases, procedures, and results – SQA Sub-team
- Update bi-directional traceability matrix (HLRs ↔ LLRs ↔ object code ↔ tests) – Documentation Sub-team
- Update certification documents – Documentation Sub-team
- Execute configuration management activities – Configuration Management Sub-team
- Complete SQA checks and audits – SQA Sub-team

This structured task decomposition provided a disciplined mechanism for integrating Agile task management practices with the stringent safety and certification requirements of DO-178C. It ensured that each user story progressed through a clearly defined, traceable, and auditable workflow, thereby harmonizing iterative Agile development with the rigor of avionics certification processes. The RACI chart outlining the responsibilities of each team for the aforementioned tasks is provided in Table 1.

Table 1: RACI chart outlining the responsibilities of each sub-team (Legend: R = Responsible-does the work, A = Accountable-owns the outcome / decision-maker, C = Consulted-provides input / feedback, I = Informed-kept updated)

| Task | Software Development | Analysis & Testing | Simulation Software Development | Documentation | Configuration Management | Certification | SQA |
|---|---|---|---|---|---|---|---|
| Develop LLRs | R/A | C | I | I | I | I | C |
| Review & analyze LLRs | R | C | I | I | I | I | A |
| Create software design | R/A | I | I | I | I | I | C |
| Review & analyze software design | R | I | I | I | I | C | A |
| Implement Software | R/A | I | I | I | I | I | C |
| Review and analyze the code | R | I | I | I | I | C | A |
| Develop simulation software LLRs | I | C | R/A | I | I | I | C |
| Create simulation software design | I | C | R/A | I | I | I | C |
| Implement simulation software | I | C | R/A | I | I | I | C |
| Static code analysis | C | R/A | I | I | I | I | C |
| Develop & execute unit tests | C | R/A | C | I | I | I | C |
| Develop & execute integration tests | C | R/A | C | I | I | I | C |
| Develop & execute system tests | C | R/A | C | I | I | I | C |
| Review verification activities | I | C | I | I | I | C | R/A |
| Update traceability matrix | C | C | I | R | C | A | C |
| Update certification artifacts | C | C | C | R | C | A | C |
| Carry out CM activities | C | C | C | I | R/A | C | C |
| Complete SQA checks | C | C | I | I | I | C | R/A |

The Daily Scrum is a time-boxed 15-minute event held at a consistent time and location each working day of the sprint, where members of the Scrum Team synchronize their work and plan for the next 24 hours [23]. In the context of aerospace software development governed by DO-178C, this event was adapted to explicitly incorporate certification-related considerations alongside conventional updates on completed work, planned tasks, and impediments. Team members convened around the task or Kanban board [76] and reported not only on functional progress but also on activities relevant to DO-178C objectives, including requirements traceability, generation of verification evidence, and enforcement of coding standards such as MISRA C/C++. The discussion also included surfacing certification-related risks, pending reviews, and gaps in documentation that might compromise compliance. This adaptation ensured that certification remained a visible and continuous priority throughout the sprint, fostering proactive alignment between agile practices and regulatory obligations, thereby reducing the likelihood of late-stage non-conformities during certification audits or reviews.

Given the multi-disciplinary and interdependent nature of aerospace projects, the Daily Scrum also served as a vital coordination forum across multiple sub-teams with specialized roles. For instance, the analysis and testing sub-team could not initiate its activities until outputs were delivered by software development and simulation software development sub-teams; similarly, the documentation sub-team required validated data from software development and analysis and testing sub-teams to prepare certification-ready artifacts. These teams were intentionally kept separate to satisfy DO-178C's requirements for independent verification and to maintain the separation of documentation from development activities in line with Agile principles [70]. The Certification sub-team participated in every Daily Scrum, monitoring progress toward certification objectives, capturing evidence, and addressing compliance-related queries in real time. Across all Scrum events, the Daily Scrum was consistently regarded by team members as the most productive, since it enabled early identification and resolution of collaboration and compliance challenges, well before they could escalate into findings during Stage of Involvement (SOI) audits or be raised as acceptance concerns by the Product Ownership team.

Once the development team completed the Sprint Backlog items and considered the increment potentially shippable within the stipulated timeframe, the subsequent step involved conducting the Sprint Review and acceptance activities to determine conformity with the Definition of Done (DoD). Within Scrum, the Product Ownership (PO) team is responsible for validating whether the delivered increment satisfies the acceptance criteria defined for each user story in the Sprint Backlog. These criteria establish the measurable conditions under which a user story can be deemed complete. In the context of aerospace software, however, the notion of "done" extends beyond functional correctness to encompass compliance with DO-178C objectives, as well as adherence to stringent safety, reliability, and certification requirements. The PO team ensured that the implemented functionality aligns with business needs and stakeholder expectations, thereby delivering the intended operational value. Concurrently, the Certification sub-team verified that the increment conforms to approved plans, maintains bidirectional traceability, and meets applicable certification standards.

This dual validation process ensures that not only the final product but also each incremental delivery is certifiable in its own right. Consequently, certification evidence can be presented to regulatory authorities (e.g., FAA/EASA) after any Sprint, if requested, thereby reinforcing transparency and regulatory confidence in the Agile process. As an illustration, consider the user story HLR-1—Altitude Hold Mode Engagement, the associated acceptance criteria are as follows:

- The FCC correctly receives the Altitude Hold mode engagement command from the Mission Management Computer (MMC) via the defined communication interface, in accordance with the Interface Control Document (ICD) and System Design Description (SDD).
- The FCC correctly stores the current barometric altitude (target altitude) received at the time of mode engagement within non-volatile or protected memory for control reference.
- The FCC correctly validates the received engagement command and target barometric altitude for checksum, message ID, and data range.
- The FCC ensures that the barometric altitude data received from the MMC is valid, current, and within plausible operational limits before being used as the target altitude.
- The FCC engages Altitude Hold only if all preconditions are satisfied—including valid sensor inputs, healthy actuator status, and no active critical faults.
- All identified safety-related requirements are addressed, implemented, and verified.
- Derived requirements are documented, reviewed, and formally approved.
- All applicable DO-178C Level A objectives have been satisfied, and this is documented in the Compliance Matrix, which maps each objective to its corresponding evidence.
- Sufficient certification evidence is generated for inclusion in Stage of Involvement (SOI) audits.

The Sprint Retrospective provides the Scrum Team with a structured opportunity to reflect on the outcomes of the sprint, identify challenges, and consider potential improvements to processes and practices [23]. In the context of DO-178C–compliant aerospace software development, the retrospective was extended beyond conventional process improvement to systematically address certification-related considerations. Alongside discussions on collaboration, tooling, and productivity, the team evaluated the degree to which DO-178C objectives were fulfilled during the sprint. This included assessing whether verification activities were thoroughly documented, requirements traceability was maintained across all development artifacts, coding standards such as MISRA C/C++ were consistently applied, and whether sufficient certification evidence had been generated to support future Stage of Involvement (SOI) audits. By embedding such reflections, the retrospective functioned as a mechanism for identifying compliance gaps, improving coordination among software development, analysis and testing, documentation, and certification sub-teams, and refining processes to ensure that each sprint not only delivered functional increments but also produced certifiable artifacts aligned with regulatory and safety standards.

The retrospective also served as a forum to gather feedback on adaptations of Scrum for aerospace projects operating under certification constraints. For example, several team members suggested extending the Daily Scrum from the standard 15 minutes to 30 minutes, as multiple interdependent teams needed to coordinate not only on development tasks but also on certification-related activities. Furthermore, the team highlighted the potential for improving efficiency through process automation, recommending the use of tools such as Atlassian Jira [71] for backlog and workflow management, as well as the development of customized tooling that integrates certification-specific activities, preconfigured templates, and audit workflows in alignment with DO-178C requirements. Collaboration challenges were also a recurring theme, particularly in managing dependencies between development, verification, documentation, and certification activities. The retrospective reaffirmed that the Daily Scrum was an effective event for resolving such issues early and that participation by certification sub-team experts was essential. In addition, it was proposed that SOI audits should be planned and conducted in parallel with sprints to avoid project delays, ensuring that certification oversight progresses concurrently with development. To support this dual focus, it was recommended that the certification liaison function be organized into two dedicated sub-teams: one embedded within the sprint team to track certification-related progress, and another interfacing directly with regulatory authorities to manage SOI audits.

After the completion of approximately half of the Sprints, the certification sub-team coordinated with the certification authority to schedule the development review (SOI #2) and verification review (SOI #3). According to FAA Order 8110.49, Software Approval Guidelines [67], SOI #2 and SOI #3 should be conducted when a representative portion (typically at least 50%) of the software development data (requirements, design, and code) and software verification/testing data, respectively, has been completed and reviewed. In alignment with this guidance, it was mutually agreed between the certification sub-team and the certification authority to conduct these reviews once sufficient evidence had been generated to demonstrate compliance with FAA Order 8110.49. During these reviews, the certification sub-team provided a comprehensive briefing covering the system under review, the adopted software life cycle model (Scrum in this case), the processes and tools employed, and any additional considerations. The regulatory authorities were engaged by presenting examples of successful aerospace projects certified at Software Level A that had employed Agile, and specifically Scrum, thereby reinforcing confidence in the safety and robustness of the approach. It was emphasized that Scrum does not compromise safety; rather, it enhances it by focusing on software quality. As noted in prior research [9], Agile practices improve both quality and productivity by emphasizing working software over documentation—consistent with the principle that it is the software itself, not the supporting documents, that ultimately operates within the aircraft. To ensure this focus, the software development and verification teams were kept independent from the documentation and certification sub-teams, while still maintaining close collaboration in line with the core values of Agile [70]. Furthermore, DO-178C explicitly permits the use of any life cycle model, provided that all objectives are satisfied in both letter and spirit [1]. Finally, in preparation for these reviews, all required data and documents were generated in advance and presented to the certification authority in accordance with FAA Order 8110.49 [67].

During each Sprint, the Product Ownership (PO) team remains actively engaged in refining and prioritizing the Product Backlog to ensure that it is adequately prepared for subsequent iterations. When new requirements are introduced, or when modifications to existing requirements or backlog items arise as a result of Sprint activities, these changes are not incorporated informally. Instead, they are formally subjected to the established Change Control process managed by the configuration management sub-team in accordance with DO-178C objectives. This ensures that every modification undergoes systematic evaluation, traceability, and approval, thereby maintaining compliance with safety-critical development standards while preserving the agility of the Scrum process.

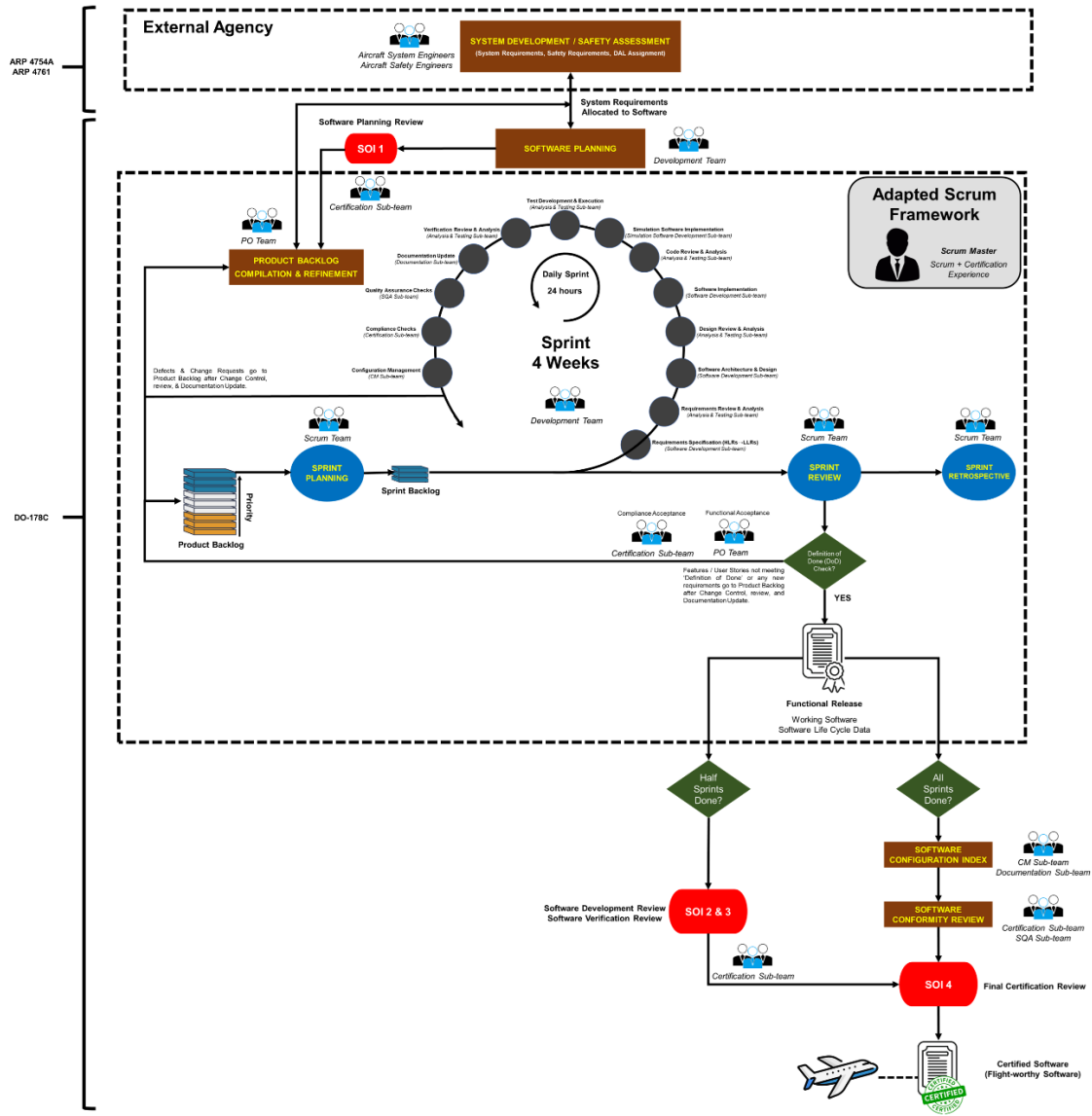

Figure 7: Adapted Scrum-based framework for DO-178C–compliant aerospace software development.

In the context of aerospace software development, where regulatory compliance and traceability are paramount, the management of burndown charts [69] and documentation of Sprint events require additional rigor beyond conventional Scrum practice. Burndown charts are systematically maintained to track the progress of backlog items against planned work, thereby providing both the development team and stakeholders with a transparent view of effort consumption, task completion, and potential deviations from the Sprint goal. Given the safety-critical nature of aerospace projects, these charts are archived as part of project records to support traceability and verification audits. Complementing this, detailed minutes of meeting (MoM) are formally captured for all Scrum ceremonies, including Sprint Planning, Sprint Reviews, and Retrospectives. The MoM serve as auditable artifacts that document decisions, identified risks, requirement clarifications, and action items, ensuring alignment with DO-178C configuration management and quality assurance objectives. Together, the disciplined management of burndown charts and MoM strengthens both project transparency and regulatory compliance, while preserving the iterative and adaptive nature of the Scrum framework. The project was executed and delivered within a total duration of seven months, comprising five Sprints of one month each.

The overall process integrates all Scrum events in their true intent, complemented by DO-178C–driven certification activities such as SOI audits. A consolidated view of this integrated framework is illustrated in Figure 7.

**5.4 Summary of Adaptations to Standard Scrum Practices**

To implement a Scrum-based Agile framework for aerospace software development, several adaptations were introduced to align with DO-178C requirements and the complexity of safety-critical systems.

*5.4.1Team Size Adaptation and Structure.*

Traditional Scrum guidelines recommend small, cross-functional teams of ten or fewer members to maximize communication efficiency and self-organization [23]. However, aerospace software development—especially at Design Assurance Level (DAL) A—requires specialized roles for verification, configuration management, and certification support that exceed standard team boundaries.

In the adapted framework, the team size was expanded to 28 members, organized as multiple synchronized Scrum sub-teams operating under a unified backlog and shared sprint cadence. Each sub-team focused on distinct functional domains, while coordination was ensured through Scrum events.

*5.4.2Multi-Disciplinary Product Ownership.*

Instead of a single Product Owner, a team of system engineers, aircraft-level engineers, and pilots collaboratively defined and prioritized backlog items. This ensured that safety, functional, and operational requirements were adequately represented and formalized through a Definition of Ready (DoR).

*5.4.3Re-Defined Scrum Master Role.*

The Scrum Master role was adapted to require dual expertise in Agile facilitation and DO-178C compliance, enabling effective alignment of iterative practices with certification obligations.

*5.4.43. Hybrid Development Teams.*

Cross-functional Scrum teams were retained, but additional compliance-focused roles—Configuration Management, SQA, Analysis and Testing, and Certification Liaison—were preserved to satisfy DO-178C's independence requirements. A dedicated Documentation Sub-team managed certification data pack.

*5.4.5 Expanded Definition of Done (DoD).*

Completion was redefined to include full compliance with DO-178C, covering requirements decomposition, design, coding (MISRA C/C++), verification, structural coverage analysis, and documentation. Each sprint increment was required to be certification-ready.

*5.4.6 Enhanced Verification and Validation (V&V).*

V&V extended beyond system-level testing to continuous reviews, traceability, structural coverage analysis, and independent verification, ensuring rigorous compliance with DO-178C objectives at all assurance levels.

*5.4.7 Modified Self-Organization.*

Teams retained autonomy within their specialties but adhered to DO-178C's independence principle, ensuring developers and testers remained organizationally distinct.

*5.4.8 Adapted Requirements Engineering.*

User stories were treated as high-level requirements (HLRs) and systematically decomposed into low-level requirements (LLRs) and software architecture artifacts before implementation, as mandated by DO-178C.

*5.4.9 Parallel Software Planning.*

Planning artifacts (SDP, SCMP, SQAP) were prepared in parallel with backlog refinement, ensuring early compliance with DO-178C and readiness for SOI #1 (Planning Audit). Documentation sub-team consolidated inputs across stakeholders.

*5.4.10 Integrated Certification Activities.*

Certification discussions and SOI reviews were embedded into Scrum events. The Certification sub-team coordinated with regulatory authorities, enabling SOI #2 and SOI #3 to align with mid-sprint progress, thus reconciling Agile iteration with regulatory oversight.

*5.4.11 Dual Acceptance Criteria.*

Each user story required dual validation: functional acceptance by the Product Ownership team and compliance acceptance by the Certification sub-team. Archived acceptance evidence ensured both agility and audit readiness.

*5.4.12 Strengthened Configuration Management and SQA.*

All changes to backlog items underwent formal Change Control, with SCMP and SQAP ensuring configuration integrity and independent quality audits, thus preserving traceability and compliance while maintaining agility.

## 6 EVALUATION OF PROPOSED SCRUM-BASED AGILE FRAMEWORK

The successful adaptation of Scrum to aerospace software development requires not only adherence to Agile values but also rigorous alignment with certification standards such as DO-178C. This section presents a systematic evaluation of the proposed Scrum-based framework, examining its effectiveness in balancing agility with regulatory compliance. The analysis is structured around the research questions defined in Section I, focusing on four key dimensions: alignment with Agile and Scrum values, satisfaction of DO-178C objectives, acceptance by certification authorities, and improvements in process efficiency. By assessing both the benefits realized and the challenges encountered during implementation, this evaluation provides critical insights into the feasibility, robustness, and scalability of the adapted framework within the highly regulated aerospace domain.

## 6.1 Assessment of the Defined Research Questions

To evaluate the effectiveness of the proposed Scrum-based Agile framework in the aerospace software domain, each of the four research questions defined in Section I was systematically examined. The following subsections present a detailed assessment of these questions, highlighting how the framework addresses agility, compliance, certification, and efficiency.

*6.1.1 RQ1 – Alignment of the Proposed Framework with Agile and Scrum Core Values.*

The foundation of Agile methods rests on Agile Manifesto [24], which articulates the mindset and principles guiding frameworks such as Scrum, XP, and Kanban. Its four core values are examined below in relation to proposed framework:

1. **Individuals and interactions over processes and tools:** Agile emphasizes people, collaboration, and communication over strict reliance on tools or formalized procedures. The proposed framework adheres to this value by actively involving end users, such as pilots, within the Product Ownership (PO) team, thereby ensuring continuous interaction with development team. Furthermore, Scrum events—including the Daily Scrum, Sprint Planning, Sprint Review, and Sprint Retrospective—serve as structured opportunities for collaboration and dialogue. Previously, such communication was limited and largely mediated through asynchronous channels such as email.
2. **Working software over comprehensive documentation:** Agile prioritizes the delivery of functional, value-generating software while still recognizing the necessity of documentation. Prior studies [10], [14], [77], [78], [79] have highlighted the tension between Agile principles and the extensive documentation mandated by aerospace standards such as DO-178C. To address this challenge, the proposed framework introduces a dedicated documentation sub-team, operating alongside all the sub-teams in the development team. The software development and analysis and testing sub-teams remain focused on software functionality and quality, while the documentation team produces compliance-related artifacts required by DO-178C [1]. These artifacts are subsequently validated by the certification sub-team for subsequent reviews and audits by the regulatory authorities. Importantly, both the documentation and certification sub-teams actively participate in Scrum events (e.g., Daily Scrum), in line with the first Agile value, ensuring transparency, clarity, and shared ownership. Documentation is thus limited to the scope explicitly required for compliance and agreed upon with regulators, preventing unnecessary overhead.
3. **Customer collaboration over contract negotiation:** Agile emphasizes continuous stakeholder engagement rather than strict reliance on predefined contracts. In aerospace projects, the customer landscape is often complex, comprising diverse stakeholder groups [6]. To operationalize this value, the proposed framework establishes a PO team that integrates multiple customer perspectives, including the aircraft-level system engineer, company system engineer, and pilot. This arrangement ensures that diverse viewpoints are represented and that stakeholder needs are directly communicated to developers, enabling ongoing collaboration throughout the project lifecycle.
4. **Responding to change over following a plan:** Agile accepts change as an inherent aspect of development, enabling teams to adapt quickly rather than adhering rigidly to pre-established plans. In aerospace projects, requirements evolve for various reasons, such as customer requests, conflicts between system architecture and requirements, technological advancements, or scope refinements [6]. The proposed framework retains core Scrum practices with minimal adaptations to accommodate these realities. For example, only well-defined and mature requirements are addressed in the initial Sprint, while ambiguous user stories are deferred for subsequent iterations. If, during implementation or acceptance, the PO team determines that a requirement does not meet the definition of "done," it is returned to the Product Backlog following a formal change control process, including configuration management, documentation updates, and review by certification authorities. Additionally, adjustments are made to Sprint scope when resource availability changes (e.g., temporary team reduction). This flexible approach embraces change while maintaining compliance with DO-178C.

While Scrum builds upon the values of Agile Manifesto, it introduces its own five core values that guide how teams collaborate and deliver results. According to Scrum Guide [23], these values are Commitment, Focus, Openness, Respect, and Courage, and their practical realization within proposed aerospace-adapted Scrum framework is described as follows:

1. **Commitment:** Team members commit to achieving the shared goals of the Scrum Team and supporting one another in delivering value. Within the proposed framework, commitment is reinforced through collective agreement on the Sprint Goal and the Definition of Done (DoD), which ensures that deliverables are not partially finished but are complete—tested, integrated, certifiable, and ready for use. The team autonomously selects Product Backlog Items (PBIs) for each Sprint, and members voluntarily take responsibility for tasks within their areas of expertise. This shared ownership strengthens accountability and commitment.
2. **Focus:** The Scrum Team directs its attention to the Sprint Goal and minimizes distractions that could compromise progress. Focus is ensured by clearly defining the "Definition of Ready" for PBIs, explicit acceptance criteria, and a well-articulated Sprint Goal. Compliance-related activities are embedded into these definitions, thereby preventing scope creep and maintaining concentration on the agreed objectives. Time-boxed Scrum events, such as the Daily Scrum, further help the team remain on track by addressing impediments swiftly.
3. **Openness:** Transparency in work, progress, and challenges is central to Scrum. In the proposed approach, openness is facilitated through the use of a Kanban/task board and daily stand-ups, which provide visibility into all tasks, including compliance-related activities. Certification and documentation sub-teams are fully integrated into Scrum events, ensuring clarity across domains. Burndown charts are maintained, and retrospectives are used to encourage candid feedback, with the Scrum Master nurturing a culture of openness.
4. **Respect:** Respect for the perspectives, skills, and contributions of all team members fosters collaboration across disciplines. The framework ensures this by valuing the expertise of cross-functional members—developers, testers, documentation specialists, and certification engineers—whose roles are critical in aerospace projects. Previous challenges, such as perceptions of unequal workload or recognition, were mitigated by the culture of openness and collaboration, which has strengthened mutual respect across domains.
5. **Courage:** Courage involves addressing difficult problems, embracing necessary changes, and upholding quality standards despite external pressures. The proposed approach demonstrates this value by empowering the team to challenge incomplete or ambiguous requirements, reject PBIs that are not ready for implementation, and escalate derived safety requirements to systems engineers for inclusion in the Product Backlog. During Daily Scrums, members openly surface risks, issues, or potential non-conformities without fear of blame. This environment of psychological safety encourages learning from mistakes and maintaining transparency, which is essential in safety-critical aerospace contexts.

The proposed framework demonstrates strong alignment with all four core values of the Agile Manifesto and five core values of Scrum, with targeted adaptations ensuring compatibility with aerospace software certification under DO-178C. Therefore, RQ1 is fully addressed, as the framework reflects both the spirit of Agile and the explicit values of Scrum.

*6.1.2 RQ2 – Assurance of Compliance with DO-178C Objectives.*

DO-178C is an objective-based standard, meaning it does not prescribe a specific development lifecycle model; any lifecycle approach is acceptable provided that all DO-178C objectives are satisfied [1], [4]. This flexibility theoretically permits the use of Agile and Scrum. However, for software at Level A (catastrophic failure condition), the objective set is highly stringent, including requirements such as independence, structural coverage, traceability, and tool qualification, all of which impose constraints on sprint execution and work organization.

In the proposed framework, complete DO-178C compliance artifacts were generated iteratively. After each sprint, both the working software and its accompanying DO-178C compliance data pack were delivered, ensuring that every increment was certifiable and deployable. Nonetheless, certain Level A objectives present friction with incremental Scrum, requiring targeted mitigations, as discussed below:

1. **Structural Coverage (MC/DC) Requirements**
   **Conflict:** DO-178C mandates Modified Condition/Decision Coverage (MC/DC) or equivalent structural coverage for Level A software [1]. Demonstrating MC/DC across an evolving codebase is challenging, as coverage must ultimately be shown on the final integrated executable, necessitating re-verification with each change. This creates churn and additional re-testing effort.
   **Mitigation:** The company employs a DO-330 qualified automated verification and testing tool suite that includes an integrated static analysis component capable of performing MC/DC analysis on the object code and generating traceability reports. This significantly reduces re-testing and re-verification overhead. Nonetheless, the long-term solution lies in achieving fully automated coverage collection and reporting integrated into the Continuous Integration (CI) pipeline.

2. **Verification Independence**
   **Conflict:** Several objectives require verification to be performed independently of development (e.g., reviews and testing) [1], [5]. Scrum's inherently collaborative and cross-functional teams can blur these boundaries, slowing in-sprint feedback if independence is not preserved.
   **Mitigation:** The proposed model ensures independence by maintaining a separate verification team, integrating independent reviews into sprint cadences, and treating independent verification tasks as dedicated backlog items with defined entry/exit criteria.

3. **Tool Qualification (DO-330)**
   **Conflict:** Agile practices emphasize automation (e.g., CI/CD, code generators, automated testing) [80], [81], [82]. If such tools replace or automate verification tasks, DO-330 requires formal qualification. The qualification burden can offset the intended agility benefits.
   **Mitigation:** The organization leverages DO-330 qualified tools (e.g., static analysis, coverage analysis, traceability, MISRA C++ compliance tools) to ensure automation aligns with certification. Further gains can be achieved by extending automation to workflow management, CI/CD, documentation, and configuration management.

4. **Requirements Decomposition and Traceability**
   **Conflict:** DO-178C requires high-level requirements (HLRs) to be decomposed into verifiable low-level requirements (LLRs), fully traceable to code and tests. Pure Scrum's practice of coding directly from user stories risks undermining traceability and compliance.
   **Mitigation:** In the proposed framework, system-level requirements are treated as epics, user stories as HLRs, and LLRs are explicitly derived during sprints. Detailed design artifacts (e.g., statecharts, sequence diagrams, pseudocode) are created alongside implementation, reviewed, and fully traced to tests and code.

5. **Architecture Stability and Refactoring**
   **Conflict:** Agile encourages frequent refactoring [83], [84], [85], [86], whereas DO-178C requires stable, documented design artifacts and traceability. Uncontrolled refactoring risks breaking compliance.
   **Mitigation:** Refactoring is governed by formal change control and impact analysis, with updates to design documentation, traceability matrices, and verification artifacts conducted each sprint. Architectural changes are treated as configuration-controlled items subject to full review and approval.

All DO-178C objectives for Level A software can be satisfied under the proposed Scrum-based framework, though they necessitate disciplined adaptations. The challenges posed by objectives such as MC/DC coverage, verification independence, tool qualification, requirements traceability, and controlled refactoring were addressed through process tailoring, role separation, automation, and strong configuration management. While Agile is not prohibited by DO-178C, Level A compliance demands stricter governance, heavier reliance on automation, and explicit compliance engineering. Accordingly, RQ2 is satisfied, as the proposed framework demonstrates full adherence to DO-178C objectives while retaining the iterative and adaptive benefits of Scrum.

*6.1.3 RQ3 – Alignment with Certification Authority Requirements.*

FAA Order 8110.49A [67] permits the use of non-sequential software life cycles, provided all DO-178C objectives are demonstrably satisfied. The findings of this study indicate that the adapted Scrum-based framework not only complies with certification authority expectations for Level A software but also introduces procedural advantages. Specifically, verification and compliance evidence—traditionally produced late in the development lifecycle—is now generated incrementally after each sprint. This enables early visibility of certification artefacts such as requirements decomposition, test results, traceability matrices, and verification records. Consequently, certification authorities can review and audit representative artefacts from the outset, rather than awaiting a single, large-scale evidence submission at project completion. This incremental transparency aligns closely with the FAA's preference for staged Stage of Involvement (SOI) reviews and mitigates the risk of late-stage nonconformities. Therefore, RQ3 is considered satisfied, subject to certain procedural considerations and adaptations discussed below.

1. **Adaptation of Sequential Certification Processes:** The certification process described in FAA Order 8110.49A follows a predominantly sequential approach. To accommodate incremental software delivery, it must be tailored to support iterative reviews without compromising traceability or independence. The proposed framework demonstrates that this adaptation is feasible when planned in collaboration with certification authorities.
2. **Handling of Requirement Changes Post-SOI:** Requirement evolution is inevitable in complex aerospace projects. To maintain compliance while enabling agility, the certification sub-team in the case study organization proactively engaged with regulators to define SOI timing and the composition of representative evidence packages (i.e., sprint-based snapshots) before major development phases commenced. This mutual agreement minimized audit surprises and ensured that late-stage changes were systematically managed within the approved process.
3. **Incremental Evidence Mapping to SOI Expectations:** Certification authorities typically require distinct review milestones (SOI #1 – #4) with baseline artefacts for planning, development, and verification. Continuous, unbounded change complicates auditability; therefore, structured baselines remain essential. The case study organization mitigated this challenge by predefining review timelines with the authority, conducting mid-project SOI reviews after representative sprints, and submitting cumulative, incremental evidence packages. Previous studies have shown that organizations can successfully meet FAA expectations by explicitly mapping incremental evidence to SOI criteria [4].

4. **Addressing Cultural and Communication Barriers:** Regulators and internal stakeholders are often more accustomed to plan-driven artefacts and sequential evidence structures. Lack of transparency or poor communication regarding incremental processes can erode regulatory confidence. To counter this, the certification sub-team and Scrum Master in the case study proactively briefed the certification authorities on the proposed Agile model, its compliance safeguards, and its traceability mechanisms prior to project initiation. This early alignment built mutual trust and acceptance of the adapted Scrum framework.

The proposed Scrum-based Agile framework aligns effectively with the expectations of certification authorities under FAA Order 8110.49A. It maintains full compliance with DO-178C objectives while enabling incremental evidence generation, early audit readiness, and greater process transparency. Successful adoption, however, depends on clear communication, pre-negotiated baselines, and proactive regulator engagement.

*6.1.4 RQ4 – Efficiency Gains in Aerospace Software Development Processes.*

To assess the efficiency and quality improvements achieved through the proposed Agile-based approach, a comparative evaluation was conducted between two completed aerospace software projects. Project A implemented multiple autopilot modes—Altitude Hold, Heading Hold, Airspeed Hold, Approach Hold, and Navigation Hold—within a Flight Control System using the proposed Scrum framework. Project B, in contrast, implemented a Ground Collision Avoidance System (GCAS) feature following a traditional Waterfall model.

Both projects were developed for the same aircraft platform at Design Assurance Level (DAL) A and exhibited comparable levels of complexity. Project B involved 25 team members, whereas Project A was staffed with 28 team members, reflecting an enhanced team composition that included one Subject Matter Expert (SME) serving as the Scrum Master, one pilot acting as a Product Owner (PO), and one aircraft systems and safety engineer contributing to the Product Ownership function. Project A was completed within a seven-month timeframe and successfully delivered in a certifiable state, satisfying all applicable DO-178C objectives. Conversely, Project B required nearly nine months to complete, with considerable delays attributed to extended requirements finalization during the early phases and significant rework and defect correction during final acceptance testing. The organization employs an integrated time-tracking system for both human and machine hours, enabling accurate estimation of total effort.

As discussed in Section IV, the company under study is a specialized aerospace subsystem development setup responsible for developing and maintaining Flight Control System components for an aircraft manufacturer. The aircraft-level system engineers provided the input in the form of System Requirements Documents (SRDs), supplemented with operational descriptions, system design descriptions, and interface control documents. These system requirements were comparable across both projects in terms of scope and implementation effort.

In Project A, 18 system requirements were decomposed into 80 High-Level Requirements (HLRs), treated as user stories under Scrum. Among the defined HLRs, 74 were functional in nature, encompassing the design and implementation of autopilot modes, control laws, interface management, data validation, and safety-related behaviors. The remaining 6 HLRs were classified as certification-oriented requirements, addressing traceability maintenance, configuration management, verification evidence generation, and incremental compilation of certification data packs to ensure continuous compliance with DO-178C Design Assurance Level (DAL) A objectives. The autopilot functionality is distributed across multiple avionics subsystems, including the Mission Management System (MMS), Flight Control System (FCS), Inertial Navigation System (INS), Instrument Landing System (ILS), Electro-Mechanical Management

System (EMMS), and Engine Control System (ECS). Consequently, not all System-Level Requirements are directly traceable to High-Level Requirements (HLRs) within the Flight Control System (FCS), as some requirements pertain to the functions or interfaces of other subsystems.

Each sprint began with backlog refinement and planning sessions, during which HLRs were further decomposed into Low-Level Requirements (LLRs), designed, implemented, and verified in accordance with DO-178C guidelines. In contrast, Project B followed the classical sequential Waterfall decomposition, consisting of 22 HLRs derived from 12 system requirements. The complete sets of High-Level Requirements (HLRs) defined for Projects A and B are provided in Appendix A and Appendix B, respectively.

Project A's execution plan included one month for product backlog preparation, five sprints of one month each, and a final month for certification audits and release. During execution, five requirement changes were introduced in Project A—none critical to system architecture—whereas Project B did not accommodate any post-baseline requirement changes. Under the traditional approach, late changes were discouraged due to the cost of re-verification and were typically deferred to later releases.

Despite comparable scope, both projects delivered different numbers of HLRs and achieved varying levels of process efficiency and product quality. To quantitatively compare performance, the following Key Performance Indicators (KPIs) were used:

1. **Total Effort per HLR (hours):** Total project effort divided by the number of HLRs [60].
2. **V&V Effort per HLR (hours):** Effort devoted to verification and validation activities per HLR [42], [60].
3. **Requirements Change (count):** Number of approved requirement changes incorporated [60].
4. **Defect Density per HLR (count)**: Number of defects identified during V&V per HLR, indicating the design and code robustness [87].
5. **Defect Leakage (count):** Defects discovered post-release, indicating test completeness.
6. **Mean Time to Detect (MTTD, days):** Average duration from defect injection to detection [88]:

$$MTTD = \frac{\sum(Detection\ Time - Injection\ Time)}{Total\ Defects}$$

7. **Mean Time to Resolve (MTTR, days):** Average duration from defect detection to resolution [88]:

$$MTTR = \frac{\sum(Resolution\ Time - Detection\ Time)}{Total\ Defects}$$

8. **Rework Effort (%):** Portion of total development effort spent correcting previously delivered artifacts [42]:

$$Rework\ Effort\ (\%) = \frac{Effort\ on\ Rework}{Total\ Effort} \times 100$$

The 1st measure of success was the team's achievement of being the first to complete the entire cycle and deliver a complete component. The comparison between both projects is shown in Table 2 for the 8 key performance indicators (KPI) listed above. For these calculations, only the 74 functional HLRs from Project A were considered.

Table 2: Comparison between Project A and Project B

| Metric | Project A | Project B |
| --- | --- | --- |
| Title | Autopilot Modes | Ground Collision Avoidance |
| Framework | Scrum | Waterfall |
| Complexity | High | High |

| Metric | Project A | Project B |
| --- | --- | --- |
| Design Assurance Level | A | A |
| No of HLRs (Functional) | 74 | 22 |
| Total Effort per HLR (hours) | 17 | 72 |
| V&V Effort per HLR (hours) | 7 | 28 |
| Requirements Changes | 5 | 0 |
| Defect Density per HLR | 0.25 | 0.55 |
| Defect Leakage | 0 | 2 |
| Mean Time to Detect (days) | 6 | 24 |
| Mean Time to Resolve (days) | 4 | 18 |
| Rework Effort (%) | 13 | 22 |

As summarized in Table 2, Project A successfully implemented a larger set of functional High-Level Requirements (HLRs)—74 compared to 22 in Project B—within a shorter overall development timeline, demonstrating a marked improvement in productivity through the adoption of the Scrum framework. The total effort per HLR in Project A was approximately 76% lower than in Project B (17 vs 72 hours), while the Verification and Validation (V&V) effort per HLR was 75% lower (7 vs 28 hours). These reductions can be attributed to the streamlined requirements elaboration process—enabled by Scrum's flexibility to initiate development with available, progressively refined requirements—along with early and continuous system integration, close coordination between development and verification teams, and the effectiveness of sprint-level reviews, which collectively minimized re-verification overhead and improved overall efficiency.

Project A successfully integrated five requirement changes, confirming its adaptability and responsiveness—capabilities typically discouraged in traditional Waterfall projects due to re-verification costs [5]. For example, during the testing of the Altitude Hold mode, excessive pitch oscillations were observed under turbulence conditions. The tolerance band was refined from ±100 ft to ±50 ft within the same sprint, demonstrating Agile's ability to incorporate stakeholder feedback without compromising schedule integrity.

In terms of quality metrics, defect density was significantly lower in Project A (0.25 vs 0.55 defects/HLR), suggesting improved design stability. Moreover, no post-release defects were reported in Project A, while two defects leaked through in Project B. MTTD and MTTR were reduced by ~75 % and ~78 %, respectively, indicating faster defect detection and rapid corrective actions enabled by continuous feedback and sprint-level reviews.

Finally, the rework effort was reduced from 22% in Project B to 13% in Project A, aligning with published findings where Agile methods typically reduce rework by 30–40% through incremental validation and early stakeholder involvement [60], [89], [90], [91].

Overall, these results substantiate that the proposed Scrum-based approach leads to measurable efficiency gains and quality improvements in aerospace software development, even within the rigorous compliance environment of DO-178C DAL A projects. Based on above, we can see that Project A is more efficient since it produces high quality software, therefore, RQ4 is satisfied. It is worth considering that this was achieved despite Project A having to contend with the learning curve associated with implementing the new Agile software development process, which included cultural challenges and the need to ramp up the team's familiarity with the process.

## 6.2 Team Satisfaction Index Calculation

A post-project team survey was conducted upon completion of the development activities. All the team members actively participated in the survey. Based on the collected responses, a Team Satisfaction Index (TSI) was computed to quantify overall morale, engagement, and perceived process effectiveness. The TSI serves as a composite performance indicator reflecting the health, cohesion, and sustainability of the development process over time. It is particularly useful for evaluating whether Agile practices, such as Scrum, enhance team well-being and productivity in high-rigor, safety-critical environments governed by standards such as DO-178C.

While no universally standardized formula exists for calculating TSI, the survey-based approach is widely adopted in contemporary literature [92]. In this approach, each team member rates a series of satisfaction dimensions on a five-point Likert scale (1 = Strongly Disagree, 5 = Strongly Agree). These dimensions capture subjective but essential aspects of team experience.

For each satisfaction factor, an average satisfaction score ($SS_i$) is computed using the following formula:

$$SS_i = \frac{\sum_{j=1}^{n} (Rating_{ij})}{n}$$

Where:

- n = number of respondents (team members)
- $Rating_{ij}$ = score given by member j for factor i (1–5)

The Team Satisfaction Index (TSI) is then calculated as the overall mean of all factor scores. This normalization expresses TSI as a percentage (0–100), facilitating comparison across projects.

$$Team\ Satisfaction\ Index\ (TSI) = \frac{Sum\ of\ all\ Ratings}{No\ of\ Questions\ or\ Factors} \times 20$$

The survey instrument, shown in Appendix C, was designed to evaluate the perceived effectiveness of Scrum practices, team collaboration, workload management, communication, and learning within an aerospace context. Participation was voluntary and responses were kept confidential to encourage openness. The results were intended exclusively for process improvement and research analysis. Average satisfaction scores for each factor are presented in Table 3.

Table 3: Average Satisfaction Score

| Factor | Avg. Rating (1–5) |
| --- | --- |
| F1: Collaboration and Communication | 4.6 |
| F2: Workload and Pace | 4.2 |
| F3: Scrum Process and Tools | 3.8 |
| F4: Technical Quality and Learning | 4.1 |
| F5: Overall Satisfaction | 4.3 |

Based on the values in Table 3, the TSI is calculated as follows:

$$Team\ Satisfaction\ Index\ (TSI) = \frac{4.6 + 4.2 + 3.8 + 4.1 + 4.3}{5} \times 20 = 84\%$$

A TSI score of 84 represents an "Excellent and sustainable" level of team satisfaction. Such a high TSI reflects robust morale, strong collaboration, and effective communication—factors that are critical for maintaining quality, compliance, and personnel retention in safety-critical aerospace software projects. The elevated satisfaction observed in Project A can be attributed to collaborative sprint planning, visible progress through reviews, and consistent feedback mechanisms, reinforcing the positive impact of structured Agile adoption under rigorous engineering constraints.

**6.3  Qualitative Findings from Team Survey**

In addition to the quantitative results presented earlier, qualitative feedback was collected in Section C of the survey in Appendix A to capture the team's perceptions regarding the benefits, challenges, and improvement opportunities arising from the adoption of the Scrum framework in a DO-178C–governed aerospace software project. Responses from all team members were analyzed thematically and are summarized below.

*6.3.1Benefits of Scrum Implementation.*

A majority of respondents expressed that the Scrum-based approach significantly improved process efficiency, product quality, and stakeholder satisfaction compared to the traditional Waterfall model.

- **Reduction in Excessive Upfront Planning:** Approximately 90% of participants noted that the earlier Waterfall approach placed disproportionate emphasis on upfront planning and documentation, often consuming more time and effort than actual software development. This "analysis paralysis" limited flexibility and responsiveness to changing requirements. In contrast, Scrum allowed teams to concentrate on iterative software delivery while still maintaining documentation necessary for regulatory compliance.
- **Improved Software Quality and Fewer Defects:** About 84% of respondents reported a measurable improvement in software quality and a reduction in post-integration defects. They attributed this to focused sprint goals, and enhanced collaboration, which helped clarify ambiguities early and address potential issues proactively.
- **Adaptability to Evolving Requirements:** 78% of participants indicated that, under the Waterfall model, requirements defined at project inception often became incomplete or outdated due to the long development cycle typical of aerospace projects. Any gaps or ambiguities propagated downstream, leading to costly rework and delays. Scrum framework, by contrast, absorbed requirement changes smoothly, preventing disruption to project progress.
- **Early Feedback and Cost Savings:** Around 70% of respondents highlighted that under Waterfall, stakeholder feedback—particularly from pilots and operational experts—was often obtained only during final testing phases, when design changes were expensive. Scrum's iterative development enabled feedback to be incorporated earlier through sprint reviews and backlog refinement, thereby reducing rework effort and associated costs.
- **Improved Integration and Reduced Risk:** 65% of respondents mentioned that delays in one subsystem under Waterfall frequently caused cascading impacts on dependent subsystems. Scrum's incremental deliveries ensured that partial software builds were available early for integration and testing, thereby mitigating schedule risks.
- **Enhanced Ownership and Accountability:** 60% of team members indicated that the use of shared task boards (e.g., Kanban) and self-tasking mechanisms fostered collective ownership and a stronger sense of responsibility, leading to better quality outcomes.
- **Improved Requirements Analysis Efficiency:** 55% of participants appreciated the inclusion of a multi-disciplinary Product Owner (PO) team, which enabled real-time requirements clarification and faster decision-making. This structure streamlined change request management while maintaining compliance with certification constraints.
- **Continuous Documentation and Compliance:** 40% of respondents stated that the DO-178C compliance process became more efficient due to continuous documentation updates aligned with sprint progress, resulting in fewer audit surprises and better preparedness for certification reviews.
- **Enhancing Productivity through Structured Agile Practices:** Approximately 30% of the survey participants indicated that the adoption of a well-defined task structure—coupled with systematic story point estimation techniques—had a notable positive impact on their overall efficiency and productivity. Participants emphasized that decomposing complex tasks into smaller, clearly scoped user stories, and assigning relative effort estimates through story points, enabled more accurate sprint planning and workload distribution. This structured approach not only facilitated better time management and prioritization of critical tasks but also contributed to the timely completion of deliverables, thereby enhancing both individual and team performance within the development cycle.
- **User-Centric Design through User Stories:** 20% of team members found that defining High-Level Requirements (HLRs) as user stories encouraged a user-focused design perspective, enhancing usability and relevance.
- **Faster Delivery and Customer Satisfaction:** 15% of respondents remarked that the Scrum approach, with its incremental and potentially certifiable software deliveries, allowed faster feature deployment—especially for critical bug fixes—thereby improving end-user satisfaction compared to the long release cycles of the Waterfall model. Respondents—particularly the pilot serving on the Product Ownership team—highlighted that the Scrum-based approach facilitated the operational deployment of latest features by accommodating controlled requirement changes within ongoing iterations. In contrast, under traditional development model, such late-stage feature enhancements were often deferred to future software releases, as change incorporation was discouraged in later phases.

*6.3.2 Challenges Encountered During Implementation.*

Despite the overall positive impact, the transition to Scrum also introduced several challenges specific to the aerospace software development domain, characterized by stringent certification and verification requirements.

- **Duplication of Verification Efforts:** 70% of participants observed that the iterative structure of Scrum led to repeated V&V efforts, such as MC/DC analysis and regression testing, in every sprint. While this improved early defect detection, it also increased workload compared to the single verification phase in Waterfall.
- **Code Refactoring Complexity:** 65% reported that due to tight coupling among aerospace software components, frequent code refactoring across iterations led to maintenance challenges and additional complexity in managing a large, evolving codebase.
- **Resource Overhead for Domain Experts:** 35% noted that the involvement of pilots and system engineers as part of the PO team, though beneficial for accuracy, imposed an operational overhead as these experts were diverted from their primary responsibilities.
- **Need for Rigorous Upfront Requirements Definition:** 25% emphasized that, despite Scrum's flexibility, rigorous upfront requirements elicitation remains essential in safety-critical aerospace projects to prevent unsafe or ambiguous implementations.
- **Certification Misalignment:** 10% of respondents expressed concern that current FAA and EASA certification processes are not yet fully compatible with iterative and incremental models. They recommended updates to regulatory guidelines to better accommodate Agile-based development.
- **Information Sharing Limitations:** 10% mentioned difficulties in sharing intermediate deliverables with customers and certification bodies, as the required documentation formats were tailored for monolithic, end-of-cycle submissions.
- **High Cost of Incremental Flight Testing:** 5% of respondents highlighted that although each sprint produced a potentially certifiable increment, full deployment required flight testing, which is costly and time-consuming. Therefore, certification audits (SOI reviews) and Supplemental Type Certification (STC) approvals remained aligned with the final release rather than each sprint. Participants suggested decoupling certification from development to improve efficiency.

*6.3.3 Suggested Process Improvements.*

Participants provided few recommendations to enhance the efficiency and compliance of Scrum within aerospace software projects:

- **Automation and Tool Integration:** 90% recommended greater automation in workflow management, traceability, and documentation generation through tools such as Atlassian Jira, automated testing frameworks, and requirements management systems capable of auto-generating traceability matrices. The adoption of Continuous Integration/Continuous Delivery (CI/CD) pipelines with well-defined entry and exit criteria was also emphasized to strengthen compliance with DO-178C objectives.
- **Dedicated Sprint for Code Refinement:** 60% proposed reserving a final sprint specifically for code maintenance, commenting, and readability improvements, ensuring long-term maintainability and audit readiness.
- **Separation of Certification from Development:** 45% suggested executing certification activities as a separate process rather than fully integrating them within Scrum iterations, to avoid slowing down development velocity while maintaining regulatory rigor.

*6.3.4 Summary of Qualitative Findings.*

Overall, the qualitative feedback indicates that the adoption of Scrum within a safety-critical aerospace environment delivered notable improvements in collaboration, quality, and responsiveness, while also exposing challenges related to certification alignment, iterative testing costs, and documentation overhead. The most significant benefits observed were enhanced team ownership, early stakeholder engagement, and incremental delivery of certifiable software artifacts. However, achieving optimal efficiency requires greater automation, improved regulatory alignment, and clearer separation of certification activities from the iterative development process.

## 7 RELATED WORK

The published literature on the application of Agile methodologies to safety-critical software systems remains largely exploratory, reflecting ongoing uncertainty among practitioners regarding effective strategies for adapting Agile principles in high-assurance contexts. A consistent theme across these studies is the recognition that Agile methods—originally designed for flexibility and rapid iteration—require contextual adaptation to align with the stringent documentation, traceability, and verification demands of safety and certification standards. Among Agile frameworks, Scrum has received the most attention in empirical and conceptual studies across several regulated domains, including the military [95], railway [97], and aerospace sectors [60], [98]. Furthermore, recent studies have introduced specialized variants such as R-Scrum [49] and Safe-Scrum [99], which aim to reconcile Agile iteration with safety and regulatory rigor.

Ribeiro et al. in "Scrum4DO178C" [60] present one of the most comprehensive adaptations of Scrum for achieving DO-178C Level A compliance in aerospace software development. Their study introduces enhancements such as rigorous backlog refinement, early integration of verification and validation (V&V), and automated documentation generation. These modifications led to measurable improvements in development efficiency and responsiveness to change while preserving regulatory compliance. However, the authors acknowledge challenges such as increased verification overhead and a strong dependency on tooling and organizational support. Compared with their work, the present study extends the understanding of these trade-offs by providing quantitative analyses of rework, defect leakage, and team satisfaction indices in a deployed aerospace project.

Aradhya [20] proposed an Agile process framework enhancing Scrum for aerospace development through three phases—preparation, development, and closure. The framework integrates code development, integration, and validation within each sprint, followed by documentation and system-level testing during closure. While this model demonstrates how incremental releases can coexist with compliance activities, Aradhya notes limitations such as the need for a more comprehensive early-phase specification and risks of architectural misalignment. Though conceptual in nature, this framework laid important groundwork for adapting Scrum to avionics. The present research builds upon this foundation by delivering empirical evidence from Level A projects, including quantitative metrics on defect density, rework effort, and team satisfaction.

Costa and Marques [17] identified barriers to Agile adoption in aeronautical software, particularly regarding DO-178C compliance, and proposed a four-phase adaptation of Scrum: Pre-game (planning, documentation, and architecture), Game (incremental development via reconfigurable sprints), Post-game (finalization and closure documentation), and review stage. While this structure provides a clear alignment between Scrum and certification stages, their study lacks implementation evidence or detailed discussion of practical integration mechanisms. Consequently, their conclusions remain largely theoretical, highlighting the need for further empirical validation.

Smith et al. [98] propose an Agile-aware assurance process for NASA's Orion spacecraft, aligning independent V&V (IV&V) activities with the Agile sprint rhythm through incremental "assurance waves." Their model prioritizes critical mission features and synchronizes verification deliverables to minimize lag in defect detection. This restructuring reduced rework and improved feedback cycles, demonstrating the feasibility of incremental assurance in a safety-critical environment. Although their work does not quantify performance through metrics such as defect density, it establishes a strong precedent for Agile–IV&V integration in high-criticality domains.

Messina et al. [95] introduce an Agile paradigm specifically tailored for mission-critical software, harmonizing Scrum's flexibility with the formal rigor of standards like DO-178C. Their results indicate reduced rework and improved adaptability, suggesting that iterative assurance processes can enhance both compliance and development responsiveness in regulated contexts.

Benedicenti et al. [96] validate Scrum's feasibility in defense software through the "Scrum4Defense" framework, which addresses hierarchical resistance and strict compliance constraints. Their findings underscore that organizational adaptation, rather than technical infeasibility, often poses greatest challenge to Agile adoption in safety-critical domains.

Myklebust et al. [97] apply Agile methods to railway software governed by the EN 50128 standard, illustrating that incremental safety cases and iterative assurance can meet certification expectations. Their findings parallel the concept of incremental evidence mapping employed in the present research for DO-178C compliance, reinforcing that iteration and assurance can coexist effectively in regulated systems.

Fitzgerald et al. [49] examined the adoption of Scrum by QUMAS Inc. in the pharmaceutical domain, demonstrating that Agile quality management processes can succeed within highly regulated environments. However, the study does not address scalability to large multi-subsystem projects or clarify how Agile artifacts directly support certification audits—gaps that this research aims to address through its aerospace-focused empirical validation.

Hanssen et al. [99] developed the "SafeScrum" method to integrate Scrum with the IEC 61508 safety standard. SafeScrum introduces separate backlogs for safety and functional requirements, reflecting their differing change frequencies. A two-year case study on a fire detection system validated the method and revealed the necessity of embedding quality assurance roles within Scrum teams. Their findings highlight both the complexity and importance of real-world evaluations for refining Agile methods in safety-critical domains.

Carbone et al. [42] proposed Safety-Enhanced Scrum, embedding safety analysis and assurance activities into every sprint to enhance traceability, transparency, and documentation quality. Their results confirm that Agile processes can meet compliance obligations under DO-178C and ISO 26262 without compromising flexibility.

McHugh et al. [93] introduced an Agile V-Model for medical device software, merging Agile iterations with the V-model's verification structure. Their approach demonstrates how incremental validation and continuous traceability can preserve compliance while improving development agility—conceptually aligned with this study's proposed framework for aerospace software under DO-178C.

Collectively, these studies establish that Agile frameworks, when systematically adapted, can achieve regulatory compliance without forfeiting flexibility or responsiveness. However, most prior works remain conceptual or limited to isolated case studies, with insufficient quantitative evaluation of their effectiveness in real-world safety-critical environments. The present research contributes uniquely by providing a comprehensive, empirically validated transformation framework for Agile adoption in the aerospace domain. It addresses transitional barriers such as organizational resistance, distributed team structures, documentation and configuration management under DO-178C, and integration of Scrum roles and ceremonies within certification-driven processes. This work offers a holistic view of Agile implementation in aerospace software, grounded in measurable performance outcomes and industry-validated evidence.

## 8 LIMITATIONS AND FUTURE WORKS

Despite the promising results obtained in this study, several limitations and validity threats must be acknowledged. These limitations highlight potential areas for refinement and opportunities for future research.

*8.1.1 Project Scope and Data Normalization.*

Although Projects A and B shared comparable objectives and development environments, the number of implemented High-Level Requirements (HLRs) differed largely. To mitigate this disparity, the collected data were normalized per contracted HLR when comparing results between the two projects. However, this normalization may not fully capture variations arising from differing requirement complexities or implementation efforts.

*8.1.2 Requirement Granularity.*

One notable limitation of this comparative analysis arises from the difference in the granularity and structure of requirements between the two projects. Project A, which followed the Scrum-based approach, employed a more detailed and iterative requirements elaboration process, resulting in a greater number of High-Level Requirements (HLRs) compared to Project B, which was developed using the traditional Waterfall model. The Scrum framework inherently encourages the decomposition of system requirements into smaller, testable, and traceable user stories to support incremental development and continuous verification. Consequently, Project A included 80 HLRs in total—74 functional HLRs related to system behavior and 6 non-functional HLRs pertaining to certification and process compliance.

To ensure a fair and meaningful comparison between the two projects, only the 74 functional HLRs were considered in the quantitative analysis of effort, productivity, and quality metrics. While this adjustment mitigates bias due to differing requirement decomposition levels, it may still introduce a degree of variability when comparing metrics such as effort per HLR or defect density. This limitation should therefore be recognized when interpreting the results, as the observed efficiency gains in Project A could partially reflect differences in requirements structuring rather than purely methodological improvements.

*8.1.3 Contextual Boundaries.*

The research was conducted in a company currently maintaining an avionics flight control subsystem that has already been developed, certified, and deployed. Consequently, this study primarily focuses on the integration and maintenance phase, rather than a full lifecycle development that includes concurrent hardware design, verification, and certification. Therefore, the findings may vary when applied to projects involving parallel hardware–software development or greenfield system design. The applicability of the proposed approach to less complex systems or larger multi-team organizational settings thus requires further empirical validation.

*8.1.4 Criticality Level Considerations.*

The implementation under study pertains to a subsystem categorized as DO-178C Level A, representing the highest safety-critical classification. As the rigor of certification processes and verification activities decreases for Levels B through E, it is expected that the relative benefits and challenges of adopting Scrum may differ. Future studies could investigate how the proposed framework scales across varying criticality levels and certification requirements.

*8.1.5 Team Experience and Organizational Culture.*

Further limitation lies in team composition and maturity. Most developers and engineers involved in the study had limited prior exposure to Agile and Scrum; their strengths lay primarily in aerospace domain and traditional software engineering practices. In organizations with higher Agile maturity or greater openness to change, observed results—particularly regarding productivity, collaboration, and quality—may differ significantly. This contextual dependency underscores the need to assess how organizational culture and readiness affect Agile adoption in regulated domains.

*8.1.6 Generalizability and Future Extensions.*

While several strategies were employed to enhance the robustness and generalizability of findings—including detailed documentation of scope, process adaptation, and evaluation procedures—further research is essential to validate the conclusions across diverse industrial and project contexts. The current study serves as a foundation for broader investigations into Agile–DO-178C integration and can be extended to projects involving third-party components, legacy systems, or hybrid certification environments.

*8.1.7 Confidentiality Constraints.*

Due to organizational confidentiality agreements and the sensitive nature of aerospace programs, the name of the participating company, as well as specific project identifiers, aircraft variants, and system details, could not be disclosed. This restriction was necessary to comply with non-disclosure obligations and to protect proprietary design and certification information. While such constraints ensured adherence to ethical and contractual requirements, they also limit the transparency and reproducibility of certain aspects of this study. The absence of detailed contextual information—such as system architecture, operational scenarios, and certification documentation—may restrict external researchers' ability to fully replicate or independently verify the reported findings. Nonetheless, all data relevant to the research objectives were analyzed and presented in an anonymized manner to maintain both scientific integrity and industrial confidentiality.

*8.1.8 Future Research Directions.*

Building on these findings, future work should explore:

- **Cross-Domain Application:** Evaluating the applicability of the proposed approach in other safety-critical domains such as automotive (ISO 26262) and medical (IEC 62304) systems.
- **System-Level Integration:** Investigating scalability to full system-level development involving parallel hardware–software integration and joint verification activities.
- **Automation and Toolchain Optimization:** Assessing the impact of automated verification, testing, and coverage tools on reducing manual effort and further improving efficiency under DO-178C constraints.
- **Alternative Agile Frameworks:** Examining how other Agile methodologies—such as Kanban, Extreme Programming (XP), or the Scaled Agile Framework (SAFe)—can be adapted for aerospace or similar high-assurance software environments.
- **Longitudinal Studies:** Conducting long-term empirical studies to measure sustained process efficiency, rework reduction, and certification compliance outcomes.

In conclusion, while the proposed approach demonstrates the feasibility of integrating Scrum practices into safety-critical aerospace software development, its generalization across diverse project types, criticality levels, and organizational contexts remains an open area for continued exploration.

# 9 CONCLUSION

This study confirms the feasibility and benefits of integrating Scrum practices within the DO-178C–governed aerospace software development lifecycle. Through targeted adaptations—such as independent verification roles, certification liaison integration, traceability-driven requirements decomposition, and continuous compliance evidence generation—the proposed framework achieved full compliance with Level A DO-178C objectives without compromising agility. By embedding regulatory requirements into the Definition of Done and aligning sprint deliverables with certification objectives, the approach ensured that agility was achieved without compromising rigor, thereby bridging the gap between iterative development and stringent aerospace standards. Comparative analysis against a traditional development approach demonstrated tangible efficiency gains, reduced rework, and enhanced team satisfaction, highlighting the potential of iterative assurance models in regulated environments.

However, several limitations were identified. The evaluation was conducted on a single organizational context involving maintenance-phase development of a flight control subsystem. Broader validation across multiple organizations, criticality levels, and lifecycle stages—including concurrent hardware–software development—is needed to generalize results. Future work should extend this framework to other safety-critical domains (e.g., automotive and medical), explore automation-driven verification pipelines, and assess long-term sustainability through longitudinal studies.

Overall, the research establishes that Agile methods, when systematically adapted and supported by qualified tools and disciplined governance, can successfully coexist with the stringent demands of safety-critical certification. This work thus bridges the longstanding gap between agility and assurance, laying the foundation for next-generation aerospace software engineering practices that are both adaptive and certifiable.

# A SYSTEM-LEVEL AND HIGH-LEVEL REQUIREMENTS FOR AUTOPILOT MODES—PROJECT A

This appendix presents the System-Level Requirements and the corresponding High-Level Software Requirements (HLRs) defined for the implementation of the autopilot modes—Altitude Hold, Heading Hold, Navigation Hold, Airspeed Hold, and Approach Hold. The autopilot functionality is distributed across multiple avionics subsystems, including the Mission Management System (MMS), Flight Control System (FCS), Inertial Navigation System (INS), Electro-Mechanical Management System (EMMS), and Engine Control System (ECS). Consequently, not all System-Level Requirements are directly traceable to High-Level Requirements (HLRs) within the Flight Control System (FCS), as some requirements pertain to the functions or interfaces of other subsystems. The System-Level Requirements are typically supported by design artifacts such as Pilot Operating Procedures (POPs), System Design Descriptions (SDDs), and Interface Control Documents (ICDs), which specify inter-system data exchange protocols, communication mechanisms, and operational logic. Within the Agile framework adopted for this study, the High-Level Software Requirements (HLRs) were expressed as Scrum User Stories and managed as Product Backlog Items (PBIs) to support iterative development and traceability in compliance with DO-178C.

## A.1 System-Level Requirements (SRS)

1. **SRS-AP-01—Primary Autopilot Function.** The Autopilot System (APS) shall command Flight Control Computer (FCC) to maintain or achieve pilot-selected mode (altitude, heading, navigation path, airspeed, and approach) while respecting aircraft handling, structural, and performance limits.
2. **SRS-AP-02—Supported Modes.** The APS shall implement the following modes with defined behavior and priorities: Altitude Hold (ALT), Heading Hold (HDG), Navigation Hold (NAV), Airspeed Hold (CAS), and Approach Hold (APP). Modes shall be annunciated and documented.
3. **SRS-AP-03—Altitude Hold (ALT) Mode.** The APS shall maintain aircraft's altitude within ±50 ft of current altitude if Altitude Hold mode is engaged.
4. **SRS-AP-04—Heading Hold (HDG) Mode.** The APS shall maintain the aircraft's magnetic heading as selected by the pilot within ±2 degrees under nominal flight conditions when Heading Hold mode is engaged.
5. **SRS-AP-05—Navigation Hold (NAV) Mode.** The APS shall maintain the aircraft on a predefined navigation path using waypoints and course data provided by the Mission Management Computer (MMC), with a lateral deviation not exceeding ±0.1 nautical miles under nominal flight conditions.
6. **SRS-AP-06—Airspeed Hold (CAS) Mode.** The APS shall maintain the aircraft's calibrated airspeed (CAS) within ±5 knots of the selected value when Airspeed Hold mode is engaged.
7. **SRS-AP-07—Approach Hold (APP) Mode.** The APS shall maintain the aircraft on the Instrument Landing System (ILS) localizer and glideslope during approach, keeping lateral deviation within ±0.1 dots and vertical deviation within ±0.3 dots when Approach Hold mode is engaged.
8. **SRS-AP-08—Mode Engagement Preconditions.** Each mode shall engage only if required preconditions are met (valid sensors, adequate actuator status, correct pilot arming/selection). Engagement conditions per mode are defined in System Design Description and approved by system safety assessment.
9. **SRS-AP-09—Mode Disengagement Conditions.** Modes shall disengage upon defined events: pilot manual override, actuator or sensor fault, exceedance of safe flight envelope limits, or detected inability to maintain commanded state. Disengagement shall generate appropriate alerts.
10. **SRS-AP-10—Mode Priority and Arbitration.** When multiple modes are armed or requested, the APS shall resolve conflicts using a deterministic mode priority table. The priority rules are defined in System Design Description, such that approach capture and safety modes override lower priority modes.
11. **SRS-AP-11—Pilot Interface and Annunciation.** Active, armed, and failed modes shall be clearly presented on the Flight Mode Annunciator (FMA) and relevant displays. Alerts (aural/visual) shall be provided for mode changes, failures, and disengagements.

12. **SRS-AP-12—Sensor and Data Integrity.** APS shall validate integrity and plausibility of sensor inputs (ADC, INS/GPS, radio nav, pitot/static) and shall define degraded modes and behavior when inputs are invalid, inconsistent, or unavailable.
13. **SRS-AP-13—Fault Management and Safe State.** APS shall detect internal faults and transition to a safe state (disengaged autopilot) when faults compromise safety. Faults and transitions must be logged and annunciated.
14. **SRS-AP-14—Performance and Timing.** APS shall meet bounded latencies from input acquisition to command output as defined in System Design Description and Interface Control Document. Control loop update rates, annunciation latencies, and computational deadlines shall be verified as per specifications.
15. **SRS-AP-15—Communication Scheme and ICD.** APS shall comply with defined ICD and System Design Description for communication with avionics subsystems (format, timing, message rates, message validation).
16. **SRS-AP-16—Overrides and Pilot Priority.** Pilot manual inputs that exceed defined thresholds shall immediately override autopilot commands and, where appropriate, disengage autopilot with documented behavior.
17. **SRS-AP-17—Logging and Maintenance.** APS shall record mode changes, engagements/disengagements, faults, overrides, and relevant flight parameters in a maintenance log and expose this data via the maintenance interface.
18. **SRS-AP-18 —Safety and Certification Artifacts.** APS shall provide full datapack from system requirements through software requirements, design, and verification artifacts conforming to DO-178C objectives for DAL A.

## A.2  High-Level Requirements (HLRs)

*A.2.1  Altitude Hold (ALT) Mode*

1. **HLR-1—Altitude Hold Mode Engagement.** As a System Engineer, I want the Flight Control Computer (FCC) to receive and store the Altitude Hold mode engagement command, along with the current barometric altitude (target altitude) from the Mission Management Computer (MMC), so that the system automatically initiates altitude hold control actions.
   Traceable to → SRS-AP-01, SRS-AP-02, SRS-AP-08
2. **HLR-2—Mode Engagement Status.** As a Pilot, I want the FCC to provide real-time status outputs to the MMC indicating whether the Altitude Hold mode is Active, Failed, or Inactive, so that I can maintain clear situational awareness of the autopilot operating state.
   Traceable to → SRS-AP-11, SRS-AP-17
3. **HLR-3—Receive Current Altitude.** As a System Engineer, I want the FCC to receive barometric altitude data from the Inertial Navigation Computer (INC) every $100 \pm 10$ milliseconds, so that the control algorithms can accurately compute altitude deviations.
   Traceable to → SRS-AP-03
4. **HLR-4—Altitude Error Computation.** As a System Engineer, I want the FCC to continuously compare the current altitude with the stored target altitude, so that pitch and throttle adjustments can be computed as required to maintain altitude stability.
   Traceable to → SRS-AP-03
5. **HLR-5—Pitch Command Generation.** As a System Engineer, I want the FCC to generate precise pitch control commands to minimize altitude error, so that the aircraft maintains altitude within ±50 feet of the target altitude under nominal conditions.
   Traceable to → SRS-AP-03

6. **HLR-6—Throttle Command Coordination.** As a System Engineer, I want the FCC to coordinate throttle control with pitch adjustments, so that altitude is maintained efficiently without exceeding flight envelope or engine operating limits.
   Traceable to → SRS-AP-03

7. **HLR-7—Feedback to Mission Computer.** As a Pilot, I want the FCC to transmit real-time pitch and throttle command data to the MMC, so that I can monitor automatic control actions and confirm stable autopilot operation.
   Traceable to → SRS-AP-11, SRS-AP-17

8. **HLR-8—Disengagement on Pilot Command.** As a Pilot, I want the FCC to process manual disengagement inputs from cockpit controls and deactivate the Altitude Hold mode with corresponding notification to the MMC, so that I can promptly regain direct control of the aircraft when desired.
   Traceable to → SRS-AP-09, SRS-AP-16

9. **HLR-9—Automatic Disengagement.** As a System Engineer, I want the FCC to automatically disengage the Altitude Hold mode if altitude cannot be maintained within specified limits, and to notify the MMC of the failure and updated engagement status, so that the aircraft can transition to a safe fallback mode.
   Traceable to → SRS-AP-09, SRS-AP-11, SRS-AP-13

10. **HLR-10—Altitude Data Failure Handling.** As a System Engineer, I want the FCC to detect invalid or missing altitude data from the INC and disengage the Altitude Hold mode, transmitting a fault notification and updated status to the MMC, so that the system maintains flight safety and avoids reliance on erroneous inputs.
    Traceable to → SRS-AP-09, SRS-AP-11, SRS-AP-12, SRS-AP-13, SRS-AP-17

11. **HLR-11—Sensor Data Failure Handling.** As a System Engineer, I want the FCC to automatically disengage Altitude Hold mode when a sensor or actuator failure is detected, and transmit fault and engagement status information to MMC, so that the aircraft can safely transition to an alternate mode and preserve flight stability.
    Traceable to → SRS-AP-09, SRS-AP-11, SRS-AP-12, SRS-AP-13, SRS-AP-17

*A.2.2  Heading Hold (HDG) Mode*

1. **HLR-12—Heading Hold Mode Engagement.** As a System Engineer, I want the FCC to receive and store the Heading Hold mode engagement command along with the current magnetic heading (target heading) from the MMC, so that the FCC can initiate the control actions required to maintain the commanded heading.
   Traceable to → SRS-AP-01, SRS-AP-02, SRS-AP-08

2. **HLR-13—Mode Engagement Status.** As a Pilot, I want the FCC to provide real-time output to the MMC indicating the engagement status of the Heading Hold mode (Active, Failed, or Inactive), so that I have clear situational awareness of the autopilot's current operating state.
   Traceable to → SRS-AP-11, SRS-AP-17

3. **HLR-14—Heading Data Acquisition.** As a System Engineer, I want the FCC to receive magnetic heading data from INC every $100 \pm 10$ milliseconds, so that it can continuously calculate heading deviation with high precision.
   Traceable to → SRS-AP-04

4. **HLR-15—Heading Error Computation.** As a System Engineer, I want the FCC to compute difference between the target heading and current aircraft heading, so that appropriate control corrections can be determined and applied.
   Traceable to → SRS-AP-04

5. **HLR-16—Roll Command Generation for Heading Correction.** As a System Engineer, I want the FCC to generate roll control commands proportional to the computed heading error, so that the aircraft smoothly aligns to and maintains the target heading within ±2 degrees under nominal flight conditions.
   Traceable to → SRS-AP-04
6. **HLR-17—Yaw Coordination.** As a System Engineer, I want the FCC to coordinate yaw control with roll commands, so that the aircraft maintains balanced and coordinated flight during heading adjustments.
   Traceable to → SRS-AP-04
7. **HLR-18—Feedback to MMC.** As a Pilot, I want the FCC to send continuous updates of heading status and control activity feedback to the MMC, so that I have real-time situational awareness of autopilot operations.
   Traceable to → SRS-AP-11, SRS-AP-17
8. **HLR-19—Manual Disengagement.** As a Pilot, I want the FCC to process my manual disengagement command to deactivate the Heading Hold mode and notify the MMC, so that I can regain direct control of aircraft when desired.
   Traceable to → SRS-AP-09, SRS-AP-16
9. **HLR-20—Automatic Disengagement on Overshoot or Failure.** As a System Engineer, I want the FCC to automatically disengage Heading Hold mode if the commanded heading cannot be maintained within ±3 degrees for more than 5 seconds, and report the fault and updated engagement status to the MMC, so that the system can transition to a safe fallback mode while preserving flight stability.
   Traceable to → SRS-AP-09, SRS-AP-11, SRS-AP-13
10. **HLR-21—Heading Data Failure Handling.** As a System Engineer, I want the FCC to automatically disengage Heading Hold mode upon detection of invalid or missing magnetic heading data from the INC, and transmit a fault notification to the MMC along with the updated engagement status, so that the aircraft can revert to a safe fallback mode and maintain flight stability.
    Traceable to → SRS-AP-09, SRS-AP-11, SRS-AP-12, SRS-AP-13, SRS-AP-17
11. **HLR-22—Sensor or Actuator Failure Handling.** As a System Engineer, I want the FCC to disengage Heading Hold mode upon detection of any sensor or actuator failure and transmit a fault notification to the MMC along with the updated engagement status, so that the aircraft can transition to a safe fallback mode and maintain flight stability.
    Traceable to → SRS-AP-09, SRS-AP-11, SRS-AP-12, SRS-AP-13, SRS-AP-17

*A.2.3  Navigation Hold (NAV) Mode*

1. **HLR-23—Navigation Hold Mode Engagement.** As a System Engineer, I want the FCC to receive and store the Navigation Hold mode engagement command along with the active flight plan data from the MMC, so that the FCC can initiate the control actions required to follow the designated navigation route..
   Traceable to → SRS-AP-01, SRS-AP-02, SRS-AP-08
2. **HLR-24—Mode Engagement Status.** As a Pilot, I want the FCC to provide real-time output to the MMC indicating the engagement status of the Navigation Hold mode (Active, Failed, or Inactive), so that I have clear situational awareness of the autopilot's current operating state.
   Traceable to → SRS-AP-11, SRS-AP-17
3. **HLR-25—Receive Navigation Data.** As a System Engineer, I want the FCC to receive navigation data from the MMC—including next waypoint coordinates and desired heading—along with current magnetic heading and coordinates data from the Inertial Navigation Computer (INC) every 100 ± 10 milliseconds, so that lateral deviation can be computed accurately.

Traceable to → SRS-AP-05

4. **HLR-26—Lateral Deviation Computation.** As a System Engineer, I want the FCC to continuously compare the aircraft's current position with the desired navigation path, so that it can determine the lateral error and compute the necessary roll or yaw control corrections.
   Traceable to → SRS-AP-05

5. **HLR-27—Roll Command Generation.** As a System Engineer, I want the FCC to generate roll control commands proportional to the computed lateral deviation, so that the aircraft remains aligned with the intended navigation path within ±0.1 nautical miles under nominal conditions.
   Traceable to → SRS-AP-05

6. **HLR-28—Yaw Coordination and Stability.** As a System Engineer, I want the FCC to coordinate yaw control with roll commands, so that aircraft maintains coordinated turns and lateral stability while following the navigation route.
   Traceable to → SRS-AP-05

7. **HLR-29—Feedback to Mission Computer.** As a Pilot, I want the FCC to send roll and yaw command data, along with cross-track deviation information, to the MMC, so that I can maintain situational awareness of the aircraft's navigation performance and autopilot actions.
   Traceable to → SRS-AP-11, SRS-AP-17

8. **HLR-30—Manual Disengagement.** As a Pilot, I want the FCC to process my manual disengagement command to deactivate Navigation Hold mode and notify MMC, so that I can regain direct control of the aircraft when desired.
   Traceable to → SRS-AP-09, SRS-AP-16

9. **HLR-31—Automatic Disengagement on Excessive Deviation.** As a System Engineer, I want the FCC to automatically disengage Navigation Hold mode when the lateral deviation exceeds ±0.3 nautical miles for more than 5 seconds, and transmit a fault notification and updated engagement status to the MMC, so that the aircraft can transition to a safe fallback mode while maintaining flight stability.
   Traceable to → SRS-AP-09, SRS-AP-11, SRS-AP-13

10. **HLR-32—Navigation Data Failure Handling.** As a System Engineer, I want the FCC to automatically disengage Navigation Hold mode upon detection of invalid or missing navigation data from the MMC or the INC, and transmit a fault notification along with updated engagement status to the MMC, so that the aircraft can revert to a safe fallback mode and maintain stable flight.
    Traceable to → SRS-AP-09, SRS-AP-11, SRS-AP-12, SRS-AP-13, SRS-AP-17

11. **HLR-33—Sensor or Actuator Failure Handling.** As a System Engineer, I want the FCC to automatically disengage Navigation Hold mode upon detection of any sensor or actuator failure and transmit a fault notification and updated engagement status to the MMC, so that the aircraft can transition to a safe fallback mode and maintain flight stability.
    Traceable to → SRS-AP-09, SRS-AP-11, SRS-AP-12, SRS-AP-13, SRS-AP-17

*A.2.4 Airspeed Hold (CAS) Mode*

1. **HLR-34—Airspeed Hold Mode Engagement.** As a System Engineer, I want the FCC to receive the Airspeed Hold engagement command and target calibrated airspeed (CAS) value from the MMC, so that the FCC can initiate control actions required to maintain the commanded airspeed.
   Traceable to → SRS-AP-01, SRS-AP-02, SRS-AP-08

2. **HLR-35—Mode Engagement Status.** As a Pilot, I want the FCC to provide real-time output to the MMC indicating the engagement status of the Airspeed Hold mode (Active, Failed, or Inactive), so that I have clear situational awareness of the system's current operating state.
   Traceable to → SRS-AP-11, SRS-AP-17

3. **HLR-36—Airspeed Data Acquisition.** As a System Engineer, I want the FCC to receive calibrated airspeed data from the INC every $100 \pm 10$ milliseconds, so that accurate and continuous airspeed monitoring can be ensured.
   Traceable to → SRS-AP-06

4. **HLR-37—Airspeed Error Computation.** As a System Engineer, I want the FCC to calculate the deviation between the commanded and current airspeed, so that the required throttle and pitch corrections can be determined to maintain target speed.
   Traceable to → SRS-AP-06

5. **HLR-38—Throttle Command Generation.** As a System Engineer, I want the FCC to generate throttle control commands proportional to the computed airspeed error, so that the aircraft maintains the commanded airspeed within $\pm 5$ knots under nominal flight conditions.
   Traceable to → SRS-AP-06

6. **HLR-39—Pitch Adjustment Coordination.** As a System Engineer, I want the FCC to coordinate pitch control adjustments with throttle commands, so that the aircraft maintains airspeed efficiently without inducing altitude or attitude instability.
   Traceable to → SRS-AP-06

7. **HLR-40—Feedback to MMC.** As a Pilot, I want the FCC to send real-time airspeed status and throttle command data to the MMC, so that I can continuously monitor automatic airspeed control and system response.
   Traceable to → SRS-AP-11, SRS-AP-17

8. **HLR-41—Manual Disengagement.** As a Pilot, I want to manually disengage the Airspeed Hold mode with an input command to the FCC, so that I can regain direct throttle or pitch control whenever required.
   Traceable to → SRS-AP-09, SRS-AP-16

9. **HLR-42—Automatic Disengagement on Limit Exceedance.** As a System Engineer, I want the FCC to automatically disengage Airspeed Hold mode if the commanded airspeed cannot be maintained within $\pm 10$ knots for more than 5 seconds, and to notify the MMC of the disengagement event and updated mode status, so that the aircraft can revert to a safe fallback state.
   Traceable to → SRS-AP-09, SRS-AP-11, SRS-AP-13

10. **HLR-43—Airspeed Data Failure Handling.** As a System Engineer, I want the FCC to detect invalid or missing airspeed data from the INC, automatically disengage Airspeed Hold mode, and report the fault and updated engagement status to the MMC, so that system safety and situational awareness are maintained.
    Traceable to → SRS-AP-09, SRS-AP-11, SRS-AP-12, SRS-AP-13, SRS-AP-17

11. **HLR-44—Sensor or Actuator Failure Handling.** As a System Engineer, I want the FCC to automatically disengage Airspeed Hold mode upon detection of any throttle actuator or control surface failure, and notify the MMC of the fault and disengagement, so that the aircraft can transition to a safe fallback mode or manual control.
    Traceable to → SRS-AP-09, SRS-AP-11, SRS-AP-12, SRS-AP-13, SRS-AP-17

*A.2.5  Approach Hold (APP) Mode*

1. **HLR-45—Approach Hold Mode Engagement.** As a System Engineer, I want the FCC to receive and store the Approach Hold mode engagement command from the MMC along with the active ILS frequency and course data, so that the aircraft can automatically track the ILS localizer and glideslope during approach.
   Traceable to → SRS-AP-01, SRS-AP-02, SRS-AP-08

2. **HLR-46—Mode Engagement Status.** As a Pilot, I want the FCC to transmit the Approach Hold mode engagement status (Active, Failed, or Inactive) to the MMC, so that I can confirm and monitor the autopilot's operating state during approach.
   Traceable to → SRS-AP-11, SRS-AP-17

3. **HLR-47—ILS Data Acquisition.** As a System Engineer, I want the FCC to receive localizer and glideslope deviation signals from the Instrument Landing System Receiver (ILS) every 100 ± 10 milliseconds, so that continuous monitoring of approach path alignment is ensured.
   Traceable to → SRS-AP-07

4. **HLR-48—Lateral and Vertical Deviation Computation.** As a System Engineer, I want the FCC to compute the aircraft's deviation from the ILS localizer and glideslope paths, so that corrective roll and pitch commands can be generated to maintain the correct approach trajectory.
   Traceable to → SRS-AP-07

5. **HLR-49—Roll Command Generation for Localizer Capture.** As a System Engineer, I want the FCC to generate roll control commands proportional to localizer deviation, so that the aircraft maintains precise lateral alignment with the runway centerline during approach.
   Traceable to → SRS-AP-07

6. **HLR-50—Pitch Command Generation for Glideslope Tracking.** As a System Engineer, I want the FCC to generate pitch control commands proportional to glideslope deviation, so that the aircraft maintains the proper vertical descent path during approach.
   Traceable to → SRS-AP-07

7. **HLR-51—Throttle Coordination for Descent Management.** As a System Engineer, I want the FCC to coordinate throttle commands with pitch control, so that the aircraft maintains a stable descent rate along the glideslope without exceeding acceptable sink rates or flight envelope limits.
   Traceable to → SRS-AP-07

8. **HLR-52—Feedback to MMC.** As a System Engineer, I want the FCC to transmit real-time lateral and vertical deviation data, along with control command feedback, to the MMC, so that the pilot can monitor approach alignment and autopilot performance in real time.
   Traceable to → SRS-AP-11, SRS-AP-17

9. **HLR-53—Manual Disengagement.** As a Pilot, I want to be able to manually disengage the Approach Hold mode, so that I can assume manual control of the aircraft during the final landing phase or when approach conditions require pilot intervention.
   Traceable to → SRS-AP-09, SRS-AP-16

10. **HLR-54—Automatic Disengagement on Signal Loss.** As a System Engineer, I want the FCC to automatically disengage Approach Hold mode when ILS signals become unreliable or are lost for more than 2 seconds, and notify the MMC of the updated engagement status, so that the aircraft can safely revert to manual or alternate control modes.

Traceable to → SRS-AP-09, SRS-AP-11, SRS-AP-12, SRS-AP-13, SRS-AP-17

11. **HLR-55—Sensor or Actuator Failure Handling.** As a System Engineer, I want the FCC to automatically disengage Approach Hold mode upon detection of any sensor, actuator, or ILS receiver failure, and transmit a fault notification and updated engagement status to the MMC, so that the aircraft can revert to manual control or a safe fallback mode.
Traceable to → SRS-AP-09, SRS-AP-11, SRS-AP-12, SRS-AP-13, SRS-AP-17

*A.2.6  General, Safety, Data Logging, Certification, and non-Functional Requirements*

1. **HLR-56—Structural Limit Enforcement.** As a System Engineer, I want the Flight Control Computer (FCC) to verify that all control surface commands (elevator, aileron, rudder, and throttle) generated during autopilot operation remain within predefined structural and aerodynamic limits, so that aircraft safety and airworthiness are preserved.
Traceable to → SRS-AP-01, SRS-AP-14

2. **HLR-57—Command Saturation Handling.** As a System Engineer, I want the FCC to detect when any commanded actuator deflection or rate exceeds the allowable range, so that the command can be automatically limited to the maximum permissible value to prevent overstress.
Traceable to → SRS-AP-01, SRS-AP-09, SRS-AP-16

3. **HLR-58—Overspeed and Stall Prevention.** As a System Engineer, I want the FCC to continuously monitor airspeed and angle of attack data, so that control outputs are constrained to prevent entry into stall or overspeed conditions.
Traceable to → SRS-AP-01, SRS-AP-12, SRS-AP-13

4. **HLR-59—Rate Limiting and Smooth Transition.** As a System Engineer, I want the FCC to apply rate limiting and command smoothing filters, so that abrupt transitions or oscillations are avoided during autopilot engagement and mode switching.
Traceable to → SRS-AP-01, SRS-AP-09, SRS-AP-16

5. **HLR-60—Limit Violation Reporting.** As a System Engineer, I want the FCC to log and report any structural limit violation or actuator command saturation to the MMC, so that the pilot and maintenance personnel can evaluate flight safety and system performance.
Traceable to → SRS-AP-09, SRS-AP-11, SRS-AP-17

6. **HLR-61—Operational Data Logging.** As a System Engineer, I want the FCC to record key flight parameters—including mode transitions, control commands, and sensor readings—so that post-flight performance analysis and diagnostics can be conducted.
Traceable to → SRS-AP-17

7. **HLR-62—Fault and Event Recording.** As a System Engineer, I want the FCC to automatically log all detected faults, sensor failures, or disconnections, with timestamps and error codes, so that root-cause analysis can be performed efficiently.
Traceable to → SRS-AP-13, SRS-AP-17

8. **HLR-63—Log Retention and Integrity.** As a System Engineer, I want the FCC to store operational logs in non-volatile memory using integrity validation methods (e.g., checksum or CRC), so that the data remains secure and verifiable for certification and maintenance.
Traceable to → SRS-AP-17, SRS-AP-18

9. **HLR-64—Maintenance Download Interface.** As a Maintenance Engineer, I want the FCC to provide a secure interface for downloading data logs through the Ground Test System, so that post-flight maintenance and fault analysis can be performed.
   Traceable to → SRS-AP-17

10. **HLR-65—Secure Log Access.** As a Software Security Engineer, I want the FCC to ensure that only authorized personnel can access or modify stored data logs, so that flight records and certification data integrity are protected.
    Traceable to → SRS-AP-17, SRS-AP-18

11. **HLR-66—Real-Time Task Scheduling.** As a Software Engineer, I want the FCC to execute all control, monitoring, and communication tasks according to a deterministic cyclic or priority-based real-time schedule, so that timing deadlines are consistently met.
    Traceable to → SRS-AP-14

12. **HLR-67—Timing Constraint Verification.** As a Software Engineer, I want the FCC to monitor the execution time of periodic tasks and detect any overruns, so that timing integrity and control loop stability are maintained.
    Traceable to → SRS-AP-14

13. **HLR-68—Data Freshness and Latency Control.** As a System Engineer, I want the FCC to ensure that input sensor data and output actuator commands are refreshed within defined latency limits, so that control response remains valid and timely.
    Traceable to → SRS-AP-13, SRS-AP-14

14. **HLR-69—Fault Handling on Timing Violations.** As a System Engineer, I want the FCC to detect and report any missed deadlines or timing faults to MMC, and record them as well, so that timely corrective action can be taken.
    Traceable to → SRS-AP-13, SRS-AP-14, SRS-AP-17

15. **HLR-70—ICD-Based Data Mapping.** As a System Engineer, I want the FCC to exchange all inter-system messages in accordance with the Interface Control Document (ICD) and System Design Description (SDD), so that data formats, timing, and signal definitions remain consistent across avionics subsystems.
    Traceable to → SRS-AP-15

16. **HLR-71—Data Validation and Range Checking.** As a System Engineer, I want the FCC to validate all incoming data fields for format, completeness, and acceptable value ranges, so that erroneous or corrupted inputs do not affect control logic.
    Traceable to → SRS-AP-12, SRS-AP-15

17. **HLR-72—Communication Protocol Compliance.** As a System Engineer, I want the FCC to implement data communication according to the ICD-defined protocol (e.g., MIL-STD-1553B or RS-422), so that data exchange remains deterministic and reliable.
    Traceable to → SRS-AP-15

18. **HLR-73—Missing or Delayed Data Handling.** As a System Engineer, I want the FCC to detect missing or delayed data frames and retain the last valid sample or switch to a safe fallback mode, so that control continuity and safety are maintained.
    Traceable to → SRS-AP-12, SRS-AP-15

19. **HLR-74—Interface Error Reporting.** As a System Engineer, I want the FCC to log and report any interface-level error (e.g., checksum failure, timeout, or invalid format) to the MMC, so that communication integrity can be monitored and maintained.
    Traceable to → SRS-AP-11, SRS-AP-15

20. **HLR-75—Configuration Traceability.** As a Configuration Management Specialist, I want all interface variables, message identifiers, and configuration parameters to remain traceable to the current ICD and software baselines for each sprint increment, so that configuration consistency and certification traceability are continuously maintained throughout development.
    Traceable to → SRS-AP-18

21. **HLR-76—Requirements Traceability.** As a Certification Liaison Engineer, I want the Autopilot Software to maintain full bidirectional traceability between system requirements, software requirements, design elements, code, and verification results for every completed sprint, so that DO-178C traceability evidence remains up-to-date and incrementally auditable.
    Traceable to → SRS-AP-18

22. **HLR-77—Verification Evidence Generation.** As a Software Quality Assurance (SQA) Engineer, I want the system to automatically generate, organize, and store verification artifacts—including test cases, test results, code coverage, and reviews—after every sprint, so that each increment contributes to a progressively complete DO-178C verification data pack.
    Traceable to → SRS-AP-18

23. **HLR-78—Configuration and Version Control.** As a Configuration Management Specialist, I want all certification artifacts produced during each sprint (requirements, design, code, and verification results) to be version-controlled under formal baselines, so that every increment's certification evidence can be traced to its corresponding build and approval record.
    Traceable to → SRS-AP-18

24. **HLR-79—Certification Data Pack Compilation.** As the Documentation Team, I want to compile and update the certification data pack—including the PSAC, SCI, SAS, and traceability matrices—at the end of each sprint or release increment, so that the software remains continuously ready for formal review or submission at any development stage.
    Traceable to → SRS-AP-18

25. **HLR-80—Audit Readiness.** As the Certification Liaison Team, I want all certification artifacts to be structured, version-controlled, and accessible after every sprint, so that software is always ready for internal or authority audits (SOI reviews) without additional preparation, ensuring ongoing compliance with DO-178C DAL A objectives.
    Traceable to → SRS-AP-18

# B  SYSTEM-LEVEL AND HIGH-LEVEL REQUIREMENTS FOR GCAS—PROJECT B

This appendix presents the System-Level Requirements and the corresponding High-Level Software Requirements (HLRs) defined for the implementation of the Ground Collision Avoidance System (GCAS) feature. The GCAS functionality is inherently distributed across multiple avionics subsystems, including the Mission Management System (MMS), Flight Control System (FCS), Inertial Navigation System (INS), Electro-Mechanical Management System (EMMS), and Engine Control System (ECS). Due to this distributed architecture, not all System-Level Requirements are directly traceable to HLRs within the Flight Control System (FCS), as certain requirements correspond to functionalities or data interfaces managed by other avionics subsystems. The presented requirements collectively ensure that GCAS feature integrates seamlessly across subsystems while maintaining compliance with DO-178C objectives.

**B.1　System-Level Requirements (SRS)**

1. **SRS-GCAS-01—Primary Safety Objective.** The GCAS shall detect potential collisions between the ownship and terrain / obstacles and shall provide timely pilot alerts and, where authorized, avoidance guidance or automatic control actions to prevent CFIT.
2. **SRS-GCAS-02—Operational Modes.** GCAS shall support at least the following modes: Standby, Advisory, Warning, and Automatic Recovery (Autonomy). Mode transition rules, priorities, and pilots' override capabilities are defined in Pilot Operation Procedure (POP) and System Design Description (SDD).
3. **SRS-GCAS-03—Predictive Conflict Detection.** GCAS shall continuously compute predicted flight trajectories and time-to-impact (TTI) against terrain / obstacle data over a specified prediction horizon and classify risk levels based on defined thresholds.
4. **SRS-GCAS-04—Alerting and HMI.** For advisory and warning conditions GCAS shall present unambiguous multimodal alerts (visual on multifunction displays, and aural alerts in pilot headset) with standardized priority and escalation logic as defined in Pilot Operation Procedure (POP) and System Design Description (SDD).
5. **SRS-GCAS-05—Automatic Recovery Authorization.** Automatic control intervention shall be permitted only when predefined, certified criteria are met (e.g., imminent impact, pilot non-response, and safety policy permits) and shall be inhibited by defined pilot actions or safety conditions.
6. **SRS-GCAS-06—Data Integrity and Validation.** GCAS shall validate availability, plausibility, and integrity of all required inputs (position, altitude, attitude, airspeed, vertical speed, terrain / obstacle database currency). GCAS shall define degraded modes with explicit behavior when inputs are invalid or inconsistent.
7. **SRS-GCAS-07—Terrain / Obstacle Database Management.** GCAS shall utilize a versioned, certified terrain / obstacle database. Database currency and integrity shall be verified at initialization and periodically as required; database metadata and version shall be logged.
8. **SRS-GCAS-08—Failure Handling and Safe Degradation.** On detection of faults that could compromise GCAS safety, the system shall transition to a defined safe state, alert crew, and log the fault for maintenance. Safe state behavior must be defined per failure mode.
9. **SRS-GCAS-09—Interface Safety and Arbitration.** GCAS shall interoperate with other avionics systems and shall include arbitration rules preventing contradictory commands to FCS and ensuring pilot authority precedence.
10. **SRS-GCAS-10—Performance and Latency.** GCAS shall meet defined worst-case end-to-end latency and computational timing bounds (sensor input to alert / command) sufficient to satisfy safety margins derived from hazard analysis.
11. **SRS-GCAS-11—Logging and Maintenance.** All GCAS events, mode transitions, faults, pilot overrides and database metadata shall be logged in a standard format accessible via maintenance interfaces.
12. **SRS-GCAS-12—Certification Evidence.** GCAS shall produce a certification evidence package that includes requirements, design artifacts, verification results, traceability matrices, and CM records adequate for SOI review.

**B.2 High-Level Requirements (HLRs)**

*B.2.1 Core Functional Requirements*

1. **HLR-01—Mode Management.** The software shall implement a Mode Manager that maintains the GCAS mode (Standby/Advisory/Warning/Automatic Recovery), enforces mode transition logic according to the state diagram defined in the System Design Description (SDD), resolves conflicts, and publishes mode / state to the FCS, displays, and maintenance logs.
Traceable to → SRS-GCAS-01, SRS-GCAS-02, SRS-GCAS-08, SRS-GCAS-09

2. **HLR-02—Sensor Fusion and State Estimation.** The software shall compute required navigational state (position, velocity, vertical speed, attitude) from redundant sensor inputs (GNSS/INS/ADC) using validated fusion algorithms and shall provide integrity metrics (solution covariance, confidence) to downstream modules.
Traceable to → SRS-GCAS-06, SRS-GCAS-09, SRS-GCAS-10

3. **HLR-03—Predictive Trajectory Engine.** The software shall implement a deterministic predictive engine that simulates ownship kinematic trajectory over the 20-second prediction horizon using aircraft state, control constraints, and nominal pilot / automation inputs.
Traceable to → SRS-GCAS-01, SRS-GCAS-03, SRS-GCAS-10

4. **HLR-04—Terrain / Obstacle Intersection Logic.** The software shall compute intersections between predicted trajectories and terrain / obstacle database points within horizontal radius of 5 NM and vertical window of ±5000 ft.
Traceable to → SRS-GCAS-03, SRS-GCAS-07, SRS-GCAS-10

5. **HLR-05—Integrity and Input Validation.** The software shall validate all inputs (sensor plausibility, message sequence, database integrity) and, on detecting invalid inputs, shall enter degraded modes and alert the pilot.
Traceable to → SRS-GCAS-06, SRS-GCAS-08

6. **HLR-06—Alerting and HMI Interface.** The software shall generate and manage multimodal alerts (visual, aural), including priority, escalation, and repetition logic, and shall hand off required status to the cockpit displays per ICD.
Traceable to → SRS-GCAS-04, SRS-GCAS-09

7. **HLR-07—Command Generation and Arbitration.** When automatic recovery is authorized, the software shall generate avoidance guidance (or direct control commands to the FCS) constrained by command limits and arbitration rules to avoid conflict with other automation. All commands shall be rate- and magnitude-limited.
Traceable to → SRS-GCAS-01, SRS-GCAS-05, SRS-GCAS-09

8. **HLR-08—Logging and Evidence Generation.** The software shall produce auditable logs and compliance data pack containing traceability, verification outputs, event logs, and database metadata suitable for SOI submission.
Traceable to → SRS-GCAS-11, SRS-GCAS-12

9. **HLR-09—Configuration and Database Management.** The software shall support secure loading, version verification, and integrity checking of terrain / obstacle databases and configuration parameters; changes shall be logged and versioned under CM.
Traceable to → SRS-GCAS-07, SRS-GCAS-11, SRS-GCAS-12

10. **HLR-10—Watchdog, Timing and Performance.** The software shall monitor execution timing, detect overruns, and take predefined safe actions if timing or resources are violated; worst-case execution times (WCET) shall be documented and met.
Traceable to → SRS-GCAS-10, SRS-GCAS-08

*B.2.2 Predictive Detection and Thresholding*

1. **HLR-11—Prediction Horizon and TTI Calculation.** GCAS shall compute TTI for predicted intersections using a prediction horizon of 20 ± 2 seconds and shall supply TTI with confidence metrics for each predicted conflict.
   Traceable to → SRS-GCAS-03, SRS-GCAS-10
2. **HLR-12—Advisory / Warning Thresholds.** The software shall classify conflicts into advisory / warning levels using parameterized thresholds traceable to the system safety assessment. The thresholds are specified as: Advisory alert if TTI ≤ 15 s or vertical clearance ≥300 ft, Warning alert if TTI ≤ 10 s or vertical clearance ≥150 ft.
   Traceable to → SRS-GCAS-02, SRS-GCAS-03, SRS-GCAS-04
3. **HLR-13—Multi-Scenario Assessment.** The software shall evaluate multiple plausible pilot responses and aircraft performance scenarios (worst-case turn rate) to determine conflict severity and robustness of advisories.
   Traceable to → SRS-GCAS-03, SRS-GCAS-10

*B.2.3 Alerting and Pilot Interaction*

1. **HLR-14—Alert Content and Priority.** The software shall present clear alert messages, annunciations, and recommended pilot actions per level (Advisory vs Warning), including standardized text and audio prompt sets.
   Traceable to → SRS-GCAS-04, SRS-GCAS-02
2. **HLR-15—Pilot Override and Acknowledgement.** The software shall allow pilot inputs (acknowledge, inhibit, or override) to affect GCAS behavior per rules. Overrides and acknowledgements shall be logged with timestamps.
   Traceable to → SRS-GCAS-02, SRS-GCAS-05, SRS-GCAS-09
3. **HLR-16—Mode Annunciation.** All GCAS modes and degraded states shall be continuously annunciated on cockpit displays and any transition shall be highlighted with appropriate alerting.
   Traceable to → SRS-GCAS-02, SRS-GCAS-04, SRS-GCAS-09

*B.2.4 Automatic Recovery*

1. **HLR-17—Automatic Recovery Authorization Logic.** Automatic recovery shall only be initiated if certified criteria are met: TTI ≤ 8 seconds, predicted avoidance maneuver within aircraft limits, pilot inactive for ≥ 2 s, all sensor integrity checks pass with confidence ≥ 0.95, and database is valid.
   Traceable to → SRS-GCAS-01, SRS-GCAS-05, SRS-GCAS-06
2. **HLR-18—Avoidance Maneuver Generation.** When initiated, the software shall compute an avoidance maneuver (pitch/roll/airspeed targets or direct FCS inputs) that maximizes separation from terrain while remaining within structural and handling constraints.
   Traceable to → SRS-GCAS-01, SRS-GCAS-05, SRS-GCAS-09, SRS-GCAS-10
3. **HLR-19—Revert and Handover.** After automatic action, the system shall hand back control to pilot when safe and shall provide clear annunciation and rationale for actions taken. All actions are logged for post-flight review.
   Traceable to → SRS-GCAS-02, SRS-GCAS-05, SRS-GCAS-08

*B.2.5 Fault & Degraded Mode*

1. **HLR-20—Sensor Loss Behavior.** Upon loss or invalidation of critical inputs (GNSS outage, altimeter failure), GCAS shall enter a documented degraded mode (advisory only, inhibited automatic recovery) and notify the crew.
   Traceable to → SRS-GCAS-06, SRS-GCAS-08, SRS-GCAS-09

2. **HLR-21—Database Invalidity.** If terrain / obstacle database fails integrity checks or is out of date, GCAS shall inhibit predictive functionality and default to a safe advisory state (or inoperative) and alert crew.
   Traceable to → SRS-GCAS-07, SRS-GCAS-08, SRS-GCAS-11
3. **HLR-22—Internal Fault Management.** Software shall detect internal computation faults (checksum, exception states) and gracefully transition to safe states while logging error context for maintenance.
   Traceable to → SRS-GCAS-08, SRS-GCAS-10, SRS-GCAS-11

## C  TEAM SATISFACTION AND PROCESS FEEDBACK SURVEY QUESTIONNAIRE

This survey aims to measure the overall satisfaction of Scrum team members regarding the development process, collaboration, workload, communication, and project outcomes within aerospace software projects. All responses are confidential and will be used only for process improvement and research purposes.

**Section A: General Information**

Team Name_________________________
Project Name_________________________
Role_________________________
Experience (in years) _________________________

**Section B: Team Satisfaction (Quantitative Assessment)**

Please rate the following statements on a scale of 1 to 5:
(1 = Strongly Disagree, 2 = Disagree, 3 = Neutral, 4 = Agree, 5 = Strongly Agree)

| No. | Statement | Rating (1–5) |
|---|---|---|
| **Collaboration and Communication** | | |
| 1 | The team collaborates effectively during sprints. | ☐1 ☐2 ☐3 ☐4 ☐5 |
| 2 | Daily Scrum meetings are productive and focused. | ☐1 ☐2 ☐3 ☐4 ☐5 |
| 3 | Communication between team members and stakeholders is transparent. | ☐1 ☐2 ☐3 ☐4 ☐5 |
| **Workload and Pace** | | |
| 4 | The sprint workload is reasonable and achievable. | ☐1 ☐2 ☐3 ☐4 ☐5 |
| 5 | I have sufficient time to complete assigned tasks with quality. | ☐1 ☐2 ☐3 ☐4 ☐5 |
| **Scrum Process and Tools** | | |
| 6 | Scrum ceremonies (planning, review, retrospective) are effective. | ☐1 ☐2 ☐3 ☐4 ☐5 |
| 7 | The tools used for task management are efficient. | ☐1 ☐2 ☐3 ☐4 ☐5 |
| 8 | Sprint retrospectives lead to real improvements in the process. | ☐1 ☐2 ☐3 ☐4 ☐5 |
| **Technical Quality and Learning** | | |
| 9 | The code quality and testing practices are maintained well. | ☐1 ☐2 ☐3 ☐4 ☐5 |
| 10 | The team is encouraged to learn and adopt better engineering practices. | ☐1 ☐2 ☐3 ☐4 ☐5 |
| **Overall Satisfaction** | | |
| 11 | I am satisfied with my role and contribution to the project. | ☐1 ☐2 ☐3 ☐4 ☐5 |

| 12 | Overall, I am satisfied with the team's performance and environment. | ☐1 ☐2 ☐3 ☐4 ☐5 |

**Section C: Qualitative Feedback**

What aspects of the Scrum process work best for your team?
☐ ______________________________________

What are the main challenges you face in the current process?
☐ ______________________________________

How can the organization improve team satisfaction and process efficiency?
☐ ______________________________________

Any additional comments or suggestions:
☐ ______________________________________